\documentclass[runningheads]{llncs}
\usepackage[T1]{fontenc}
\usepackage{graphicx}
\usepackage[dvipsnames]{xcolor}
\usepackage{amsfonts}
\usepackage{amsmath}
\usepackage{amssymb}
\usepackage{booktabs}
\usepackage{longtable}
\usepackage{enumitem}
\usepackage{multirow}
\usepackage{tabularx}
\usepackage{tikz}

\usepackage{hyperref}
\hypersetup{
	colorlinks   = true, %Colours links instead of ugly boxes
	urlcolor     = black, %Colour for external hyperlinks
	linkcolor    = gray, %Colour of internal links
	citecolor   = gray %Colour of citations
}

\usepackage{xparse}
\usepackage{comment}
\usepackage{algorithm}
\usepackage{algpseudocode}
\usepackage[capitalise]{cleveref}
\usepackage{subcaption}
\usetikzlibrary{backgrounds, fit, shapes}
\pgfdeclarelayer{background}
\pgfsetlayers{background,main}
\usepackage{fontawesome}
\usepackage{color}

\usepackage{xspace}

\newcommand{\Agents}{\mathcal{K}}

\newcommand{\ActionSpace}[1]{\mathcal{A}_{#1}}

\newcommand{\JointMdR}{\mu}

\newcommand{\aState}{s}
\newcommand{\jointAction}{A}

\newcommand{\action}[1]{a_{#1}}

\newcommand{\mdr}[1]{\mu_{#1}}
\newcommand{\validCount}[3]{n_{#3} \bigl(#1, #2 \bigr)}

\newcommand{\intervention}[1]{\left[ \jointAction_{#1} \leftarrow \mdr{#1} \right]}

\newcommand{\nij}{\validCount{\aState}{\jointAction}{j}}
\newcommand{\nijMdR}{\validCount{\aState}{\intervention{i}}{j}}
\newcommand{\nii}{\validCount{\aState}{\jointAction}{i}}
\newcommand{\niiMdR}{\validCount{\aState}{\intervention{\neg i}}{i}}

\newcommand{\singleton}[1]{\{#1\}}
\newcommand{\nijMdRG}{\validCount{\aState}{\intervention{G}}{j}}

\newcommand{\FeAR}{\mathrm{FeAR}}
\newcommand{\clip}[1]{Z \Biggl(#1\Biggr)}

\newcommand{\group}[1]{\left\{#1\right\}}

\newcommand{\solo}[2]{#2\nolinebreak\leftharpoonup\nolinebreak#1}
\newcommand{\mediated}[3]{#2\nolinebreak\leftharpoonup_{#3}\nolinebreak#1}
\newcommand{\coupled}[2]{#2\nolinebreak\Leftarrow\nolinebreak#1}
\newcommand{\mediatedCoupled}[3]{#2\nolinebreak\Leftarrow_{#3}\nolinebreak#1}

\newcommand{\tier}[2]{\mathbb{T}_{#1,#2}}

\newcommand{\courteous}[1]{\phi_{#1}}
\newcommand{\candidate}[1]{\kappa_{#1}}

\newcommand{\phdot}{\makebox[0pt][l]{.}\phantom{0}}

\newcommand{\deltaAssertive}{\Delta_j^{\mathrm{Assertive}}}
\newcommand{\nAssertiveiFeAR}{n^\mathrm{Assertive}_{j,\mathrm{iFeAR}}}
\newcommand{\nAssertivegFeAR}{n^\mathrm{Assertive}_{j,\mathrm{gFeAR}}}

\newcommand{\fearRanksPlain}{\emph{iFeAR}}
\newcommand{\tierRanksPlain}{\emph{gFeAR-Tier}}
\newcommand{\shapRanksPlain}{\emph{gFeAR-Shapley}}

\newcommand{\fearRanks}{\text{\footnotesize\fearRanksPlain}\xspace}
\newcommand{\tierRanks}{\text{\footnotesize\tierRanksPlain}\xspace}
\newcommand{\shapRanks}{\text{\footnotesize\shapRanksPlain}\xspace}

\newcommand{\taufearTier}{\tau(\text{\scriptsize\fearRanksPlain,\, \tierRanksPlain})}
\newcommand{\taufearShap}{\tau(\text{\scriptsize\fearRanksPlain,\, \shapRanksPlain})}
\newcommand{\tauTierShap}{\tau(\text{\scriptsize\tierRanksPlain,\, \shapRanksPlain})}

\newcommand{\inColouredBox}[3]{
\begin{tikzpicture}
      \node[
        draw=#1,
        line width=2pt,
        inner sep=8pt
      ] (box) {
        \includegraphics[width=\linewidth,height=\linewidth]{#3}
      };
      \node[
      anchor=south east,
      font=\footnotesize,
      xshift=-2pt,
      yshift=2pt
    ] at (box.south east) {#2};
    \end{tikzpicture}
}

\begin{document}

\title{Using Feasible Action-Space Reduction by Groups to fill Causal Responsibility Gaps in Spatial Interactions}
\titlerunning{FeAR by Groups}

% \author{Vassil Guenov\thanks{These authors contributed equally.}\inst{1}\orcidID{0009-0001-8287-9199} \and
% Ashwin George\footnotemark[1]\inst{1,2}\orcidID{0009-0007-7655-0737} \and
% Arkady Zgonnikov\inst{1,2}\orcidID{0000-0002-6593-6948} \and
% David A. Abbink\inst{1,2}\orcidID{0000-0001-7778-0090} \and
% Luciano Cavalcante Siebert\inst{1,2}\orcidID{0000-0002-7531-3154}}

\author{
Ashwin George\thanks{These authors contributed equally to this work.}\inst{1,2}\orcidID{0009-0007-7655-0737} \and
Vassil Guenov$^\star$\inst{1}\orcidID{0009-0001-8287-9199} \and
Arkady Zgonnikov\inst{1,2}\orcidID{0000-0002-6593-6948} \and
David A. Abbink\inst{1,2}\orcidID{0000-0001-7778-0090} \and
Luciano Cavalcante Siebert\inst{1,2}\orcidID{0000-0002-7531-3154}
}

\authorrunning{A. George
et al.}
\institute{
Delft University of Technology, The Netherlands \and
Centre for Meaningful Human Control
\email{A.George@tudelft.nl}
}
\maketitle              % typeset the header of the contribution
\begin{abstract}
Heralding the advent of autonomous vehicles and mobile robots that interact with humans, responsibility in spatial interaction is burgeoning as a research topic.
Even though metrics of responsibility tailored to spatial interactions have been proposed, they are mostly focused on the responsibility of individual agents.
Metrics of causal responsibility focusing on individuals fail in cases of causal overdeterminism --- when many actors simultaneously cause an outcome.
To fill the gaps in causal responsibility left by individual-focused metrics, we formulate a metric for the causal responsibility of groups. 
To identify assertive agents that are causally responsible for the trajectory of an affected agent, we further formalise the types of assertive influences
and propose a tiering algorithm for systematically identifying assertive agents.
Finally, we use scenario-based simulations to illustrate the benefits of considering groups and
how the emergence of group effects vary with interaction dynamics and the proximity of agents.

\end{abstract}

%------------------------------------------------------------------ %

\section{Introduction}
\label{sec:Introduction}

Navigation of mobile robots and self-driving cars among humans, especially in multi-agent settings, faces many challenges owing to the complexity of multi-agent interactions~\cite{markkulaDefiningInteractionsConceptual2020,schwartingSocialBehaviorAutonomous2019}.
In these safety-critical interactions,
to make agents act responsibly and to attribute responsibility in case of accidents,
we need models of responsibility~\cite{papadimitriouCommonEthicalSafe2022,calvertDesigningAutomatedVehicle2023}.  
Research related to spatial interactions has focused on 'responsible' or 'responsibility-aware' navigation based on how agents yield to other agents~\cite{remyLearningResponsibilityAllocations2024,cosnerLearningResponsibilityAllocations2023,geisslingerEthicalTrajectoryPlanning2023,shalev-shwartzFormalModelSafe2018,shalev-shwartzVisionZeroProvable2019}. While these methods are not based on the formal definition of ``responsibility'', a more formal model of causal responsibility in spatial interactions was proposed based on how individual agents restrict the feasible action space of other agents~\cite{georgeFeasibleActionSpaceReduction2023a,george2023measuring}. However, all of these approaches only focus on the responsibility of individual agents. 

Formal models of group responsibility have dealt with the distributing the responsibility of collective outcomes to individual agents in more abstract scenarios with fewer actions, which do not scale to complex spatial interaction \cite{brahamDegreesCausation2009,alechina2017causality,yazdanpanahDistantGroupResponsibility2016,yazdanpanahApplyingStrategicReasoning2021,loriniLogicalAnalysisResponsibility2014}. Understanding group responsibility is important because the actions of individuals might have superadditive influence on the outcome when acting together \cite{duijfLogicalStudyMoral2023}, which complicates the tracing of responsibility to individuals. This is even more problematic in cases of causal overdetermination, where many agents simultaneously cause an outcome and no individual agent can be deemed causally responsible in isolation~\cite{brahamDegreesCausation2009}.

\begin{figure*}[tb]
    \centering
    \begin{subfigure}[b]{0.3\linewidth}
    \includegraphics[width=\linewidth]{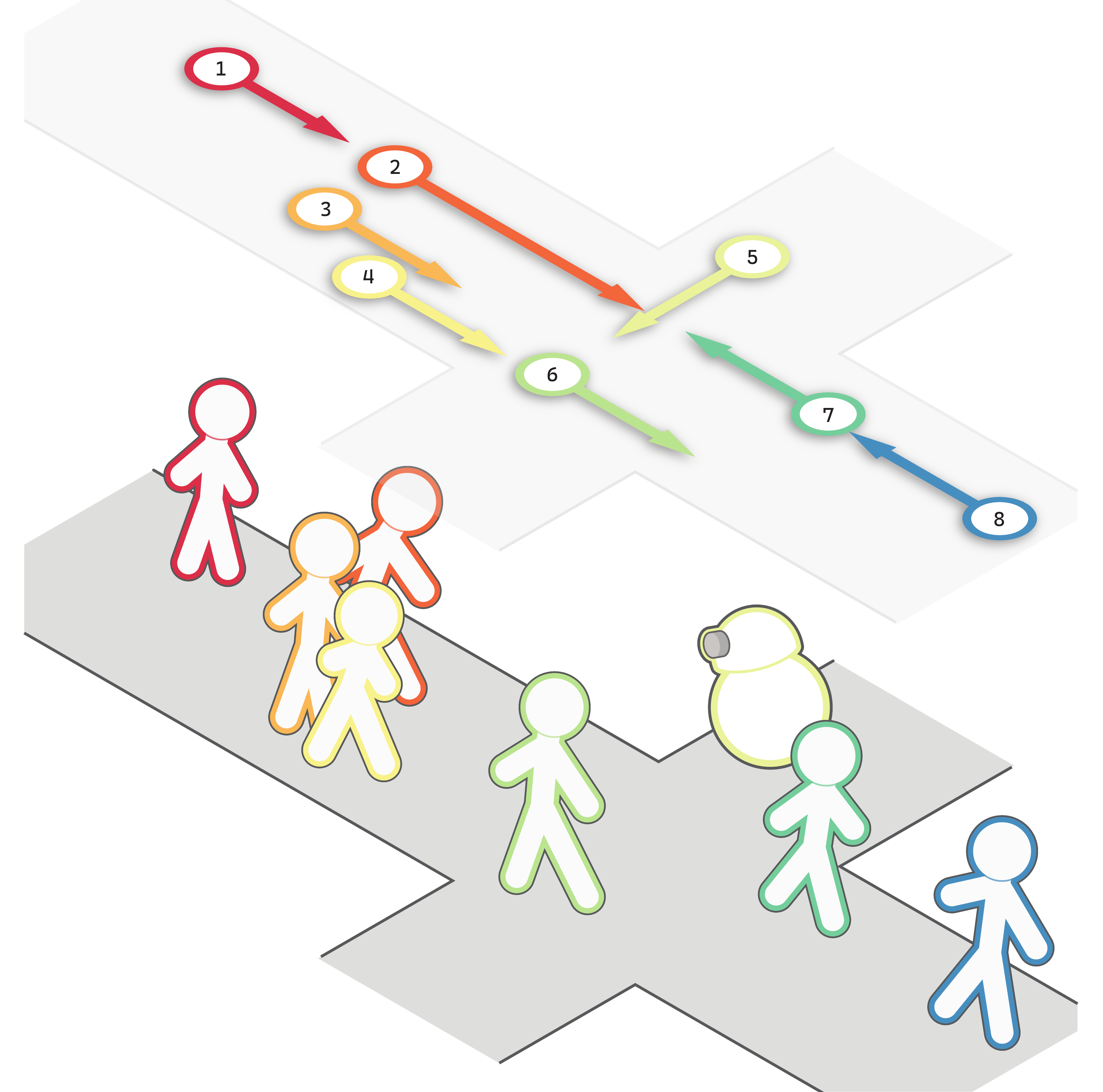}
    \caption{\footnotesize {Robot crossing}}
    \label{fig:CrossyRoadIllustration}
    \end{subfigure}
    \hfill
    \begin{subfigure}[b]{0.68\linewidth}    
    \begin{subfigure}[b]{0.48\linewidth}    
    \includegraphics[width=\linewidth]{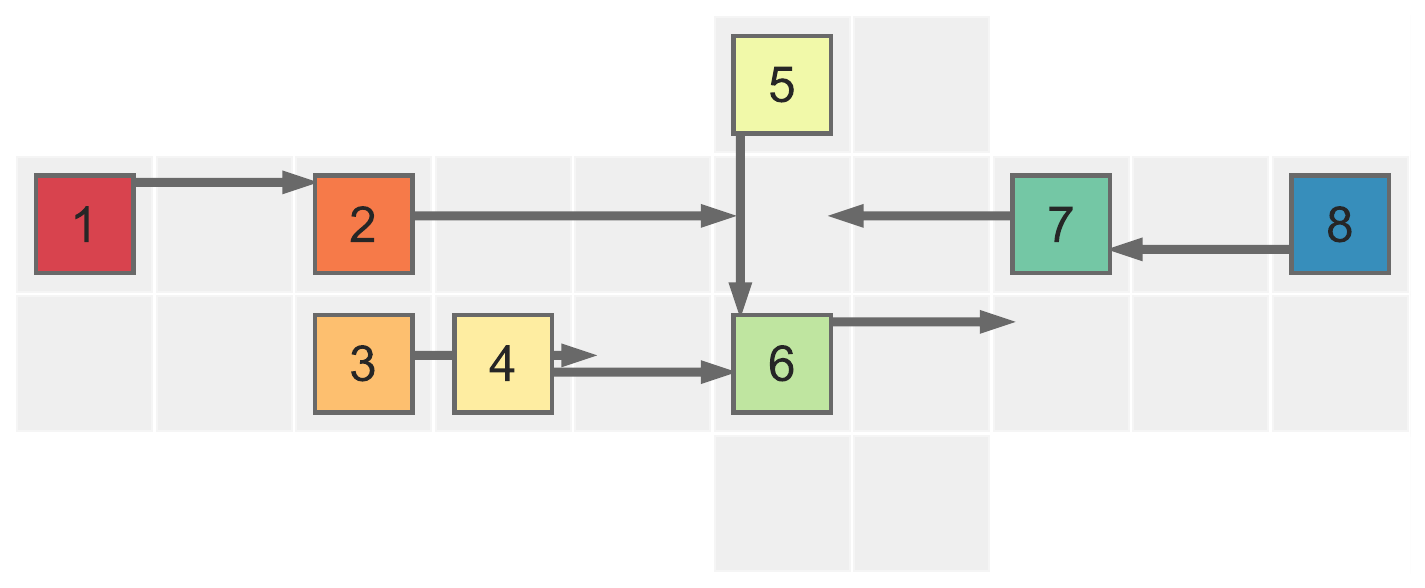}
    \caption{\footnotesize{Grid world scenario}}
    \label{fig:CrossyRoad_GWorld}
    \end{subfigure}
    \hfill
    \begin{subfigure}[b]{0.48\linewidth}    
    \includegraphics[width=\linewidth]{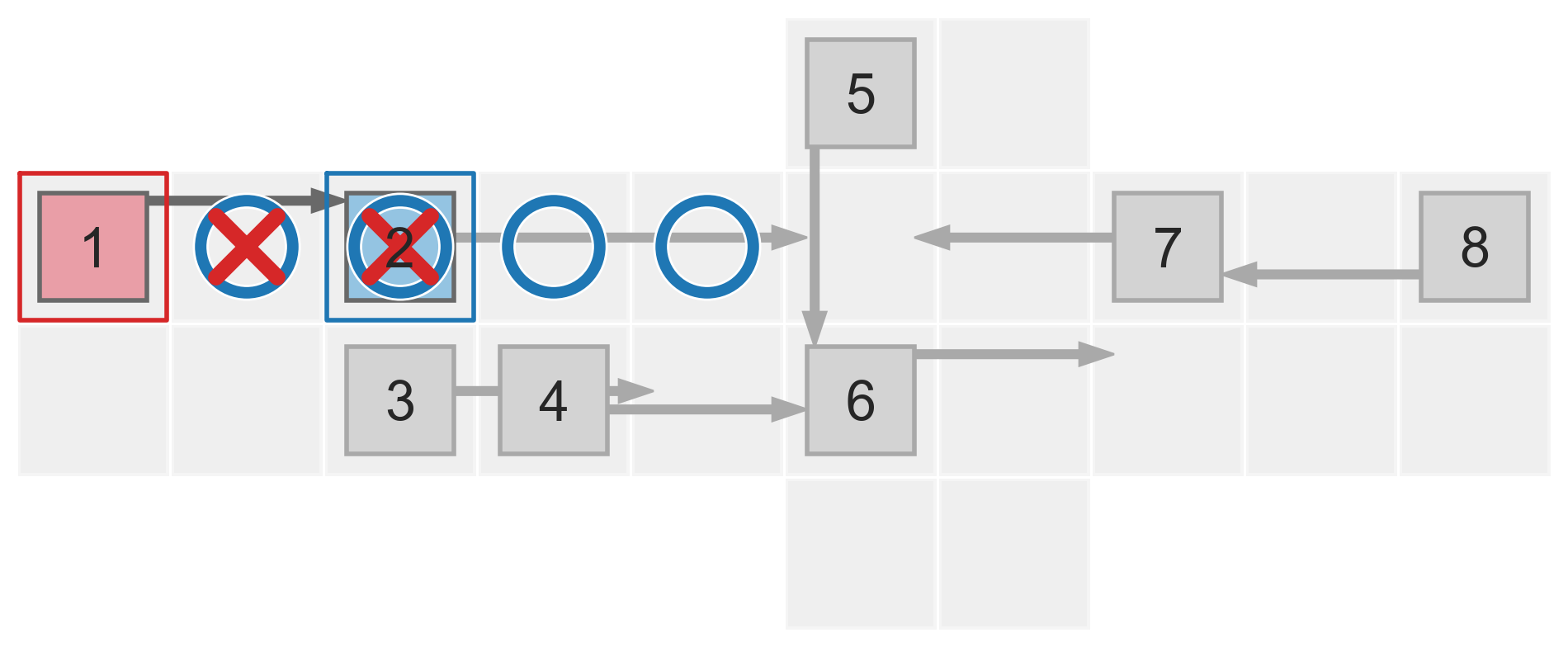}
    \caption{\footnotesize {Effect of 1 on 2}}
    \label{fig:CrossyRoad-1on2}
    \end{subfigure}
    \begin{subfigure}[b]{0.48\linewidth}    
    \includegraphics[width=\linewidth]{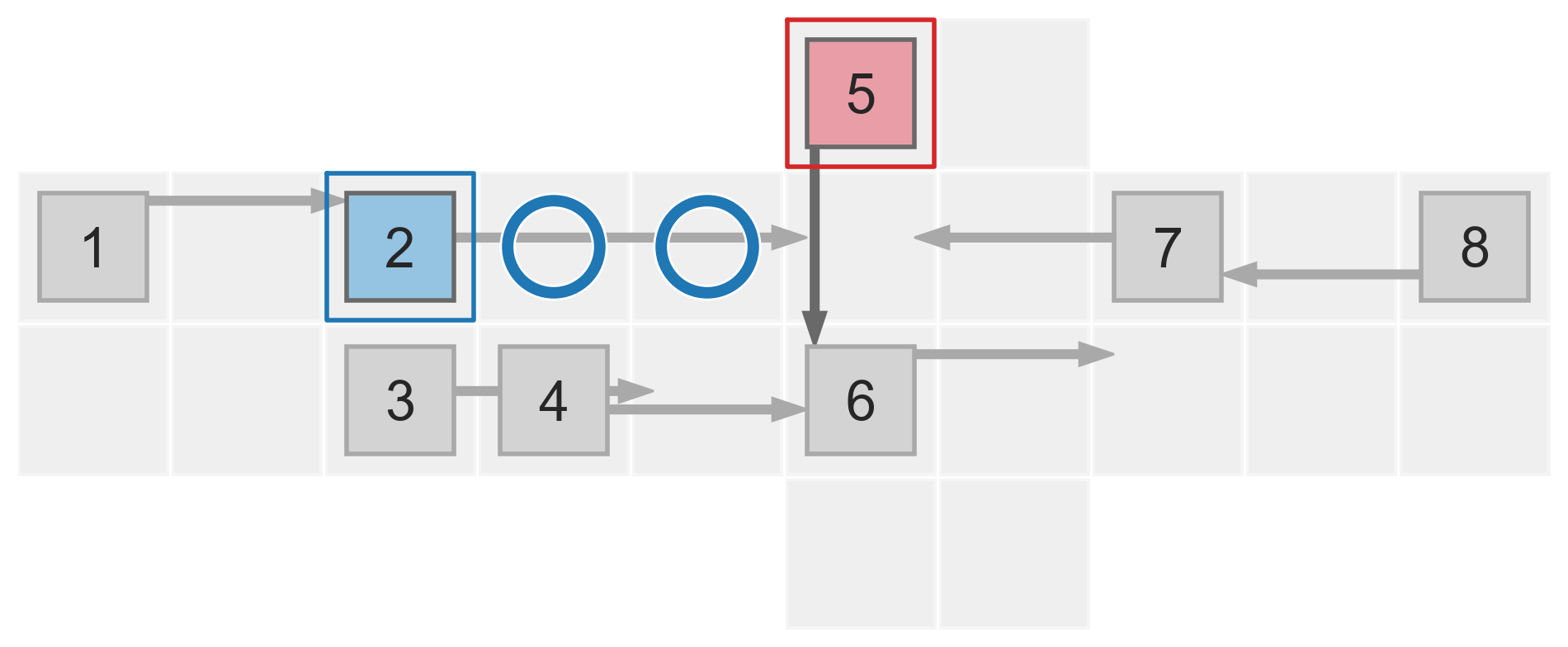}
    \caption{\footnotesize {Effect of 5 on 2}}
    \label{fig:CrossyRoad-5on2}
    \end{subfigure}
    \hfill
    \begin{subfigure}[b]{0.48\linewidth}    
    \includegraphics[width=\linewidth]{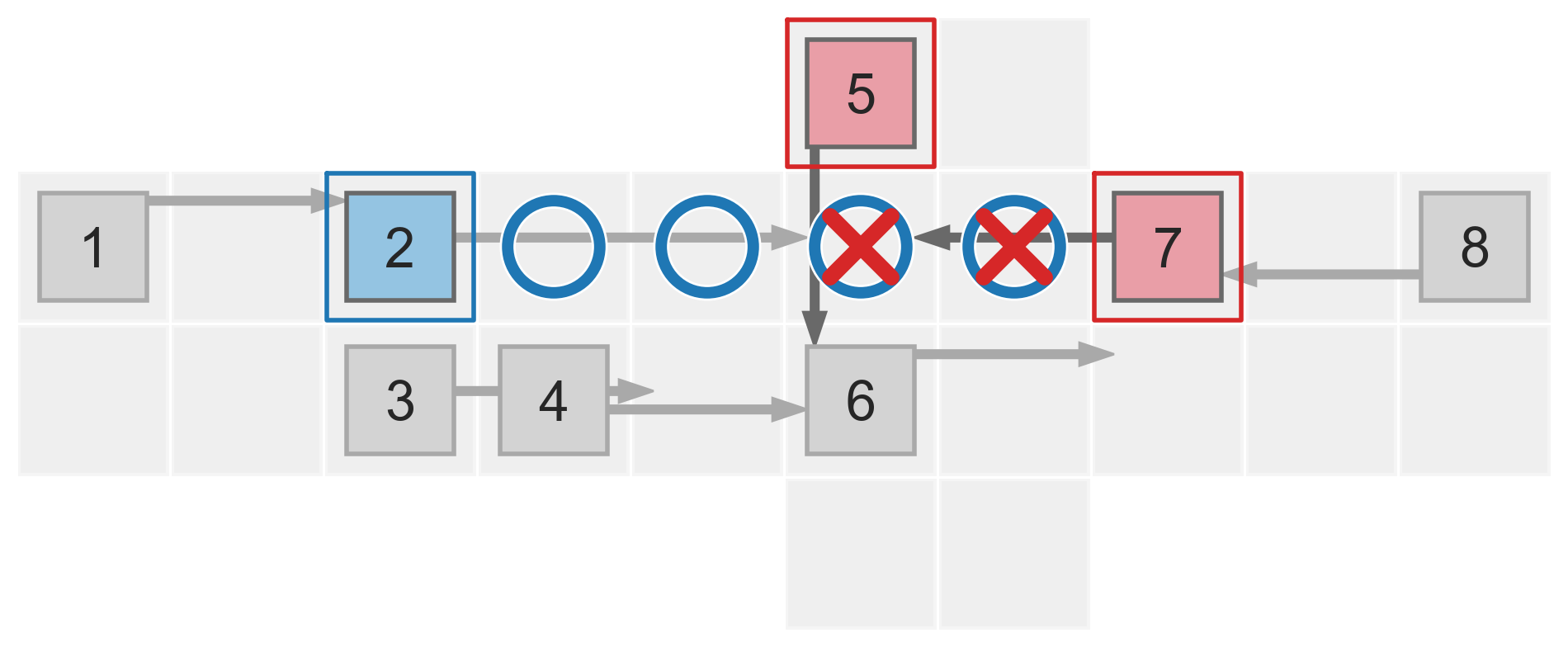}
    \caption{\footnotesize {Effect of \{5,7\} on 2}}
    \label{fig:CrossyRoad-5-7-on2}
    \end{subfigure}
    \end{subfigure}
    \caption{
    \small
    \textbf{Feasible action space reduction \cite{georgeFeasibleActionSpaceReduction2023a}:}
    For the robot crossing (a) represented in the grid world (b),
    the feasible action space reduction (FeAR) imposed by \textcolor{BrickRed}{actors} on \textcolor{RoyalBlue}{affected} agents are computed based on \textcolor{RoyalBlue}{the feasible actions of the affected when actors follow their \emph{Move de Riguer}~(MdR)} (represented by \textcolor{RoyalBlue}{\faCircleO}) and how many of these are \textcolor{BrickRed}{rendered infeasible by the actual actions of actors} (represented by \textcolor{BrickRed}{\faClose}).
    For affected agent 2, (c) shows how agent 1 reduces the feasible action space by two, (d) shows how 5 on its own has no influence, and (e) shows how the group $\group{5,7}$ reduces the feasible action space by two.}
    \label{fig:CrossyRoadIntro}
\end{figure*}

Consider the scenario in \cref{fig:CrossyRoadIntro} of a robot crossing some pedestrians, which was analysed with the Feasible Action-Space Reduction (FeAR) metric~\cite{georgeFeasibleActionSpaceReduction2023a}.
According to this metric, agents that restrict the feasible action space of another agent are causally responsible for the trajectory of the latter.
% Consider the interaction where a robot is crossing the path of some pedestrians as shown in \cref{fig:CrossyRoadIntro}.
% Following the conventions set in \cite{georgeFeasibleActionSpaceReduction2023a},
If we represent this interaction using a grid world (as in \cite{georgeFeasibleActionSpaceReduction2023a}), where the actions of agents correspond to their speed~(\cref{fig:CrossyRoad_GWorld}), we can see that the action of pedestrian 1 is restricting the feasible actions of agent 2, compared to the case when agent 1 would have stayed \cref{fig:CrossyRoad-1on2}.
Even though this insight helps us, individual FeAR fails in some cases.
For example, if you look at how agent 2 is affected by 5 and 7. Both agents 5 and 7 simultaneously restrict the same actions, and even if one of them were to stay, the feasible action space of agent 2 would not change~(\cref{fig:CrossyRoad-5on2}). But when we consider them as a group, we can see how they are collectively reducing the feasible action space of agent 2~(\cref{fig:CrossyRoad-5-7-on2}). Since individual FeAR fails in cases of causal overdetermination, relying on individual FeAR to ascribe responsibility can lead to responsibility gaps~\cite{duijfLogicalStudyMoral2023}. Responsibility gaps occur when a group of agents have collective responsibility, but no individual agent can be held responsible~\cite{duijfLogicalStudyMoral2023}.

To prevent such responsibility gaps, we reformulate the Feasible Action-Space Reduction (FeAR) metric for quantifying the causal responsibility of groups on the trajectory of an affected agent~(\cref{sec:GroupFeAR}).
For quantifying the contributions of individuals to group outcomes (as in~\cite{yazdanpanahDistantGroupResponsibility2016,yazdanpanahStrategicResponsibilityImperfect2019}), we formally categorise different types of assertive influences~(\cref{sec:TypesOfInfluence}) and use these categorisations to rank the assertive influence of different agents into tiers~(\cref{sec:RankingAgents}). These tiers were further used to explore the emergence of group effects in different scenarios~(\cref{sec:Scenarios}).

The main contributions of this paper are:
(a) a formulation of the feasible action-space reduction (FeAR) metric to quantify the causal responsibility of groups on the trajectory of other agents in spatial interactions,
(b) a formal categorisation of the types of assertive influences,
(c) a tiering algorithm for ranking the assertive influence of agents on an affected agent, and
(d) scenario-based simulations showing how the emergence of group effects are dependent on the dynamics of the interaction and proximity of the affected agent to other agents.

Rest of the work is organised as definitions~(\cref{sec:Definitions}), case studies~(\cref{sec:ScenarioSims}), results~(\cref{sec:Results}), discussion~(\cref{sec:Discussion}), and conclusion~(\cref{sec:Conclusion}).

\section{Definitions}
\label{sec:Definitions}

After some preliminary notations~(\cref{sec:Preliminaries}) and the definition of the Feasible Action-Space Reduction (FeAR) metric for individual actors~\cite{georgeFeasibleActionSpaceReduction2023a}~(\cref{sec:FeAR}), we present our formulation of FeAR for groups~(\cref{sec:GroupFeAR}). Then, we formally categorise the types of assertive influence~(\cref{sec:TypesOfInfluence}) and propose a tiering algorithm for ranking the assertive influences of agents~(\cref{sec:RankingAgents}).

\subsection{Preliminaries}
\label{sec:Preliminaries}

As proposed by \cite{georgeFeasibleActionSpaceReduction2023a}, we model
spatial interactions among a set of $k$ agents $\Agents$
using a grid world where $\aState$ represents the state of the grid world encompassing information about spatial constraints and the locations of agents. Each agent $i \in \Agents$, chooses an action $\action{i}$ which corresponds to the speed with which they move for the time window in consideration. Each agent $i$ has an action space of 17 actions, $\ActionSpace{i}$ = 
\{S0,
U1, U2, U3, U4,
D1, D2, D3, D4, 
L1, L2, L3, L4, 
R1, R2, R3, R4\}
--- where the first character indicates the direction (Stay, Up, Down, Left or Right) and the second character indicates the speed in steps moved per time window. 
$\jointAction=\left(\action{i}\right)_{i \in \Agents}$ represents the joint action of all the agents. 

We compare the actions of agents $\action{i}$ against their \emph{Move de Rigueur}(MdR) $\mdr{i}$ which represents expectations of how agents would act in a given state $\aState$. Ideally, if all agents follow the joint MdR, $\JointMdR = \left(\mdr{i}\right)_{i \in \Agents}$, there would not be any crashes.
In this paper, we consider that the MdR is staying (S0) for all the agents in all scenarios. 
The intervention of replacing the actual action with MdR is represented as follows:

\begin{equation}
\begin{aligned}    
    \intervention{G} &= \left( a'_i \right)_{i \in \Agents} \quad \text{where, }
    a'_{i} &= 
    \begin{cases}
    \mdr{i}\in \JointMdR, & \text{if } i \in G\\
    a_i \in \jointAction, & \text{if } i \notin G
    \end{cases}
\end{aligned}
\end{equation}
and, $\intervention{i} = \intervention{\singleton{i}}$ represents the intervention of replacing the action of a single agent $i$ with its MdR.
Finally, $\nij$ gives the number of feasible moves available to agent $j$ for a given state $\aState$ and joint action $\jointAction$.

\subsection{Individual FeAR}
\label{sec:FeAR}

Based on the idea that agents that restrict the feasible action space of other agents are causally responsible for the trajectory of the affected agent, the Feasible Action-Space Reduction (FeAR) metric was defined for individual actors as follows:

\begin{definition}[\textbf{FeAR}~\cite{georgeFeasibleActionSpaceReduction2023a}]

The Feasible Action-Space Reduction $(\FeAR)$ imposed by actor $i$ on affected agent $j$ is defined as:

    \begin{equation}
	\FeAR_{i,j}(\aState,\jointAction) =
	\begin{cases}
        \clip{\frac{\nijMdR-\nij}{\nijMdR + \epsilon}}, &
        \text{if } i \neq j,\\
		\clip{\frac{\nii}{\niiMdR + \epsilon}}, & \text{if } i = j,\\
	\end{cases}
     \label{Eq:FeAR}
    \end{equation}
    where $\neg i = \Agents \setminus \singleton{i}$,  $Z(x)= \min\left(-1, \max(x,1)\right)$, and 
    $0<\epsilon\ll1$.
\end{definition}

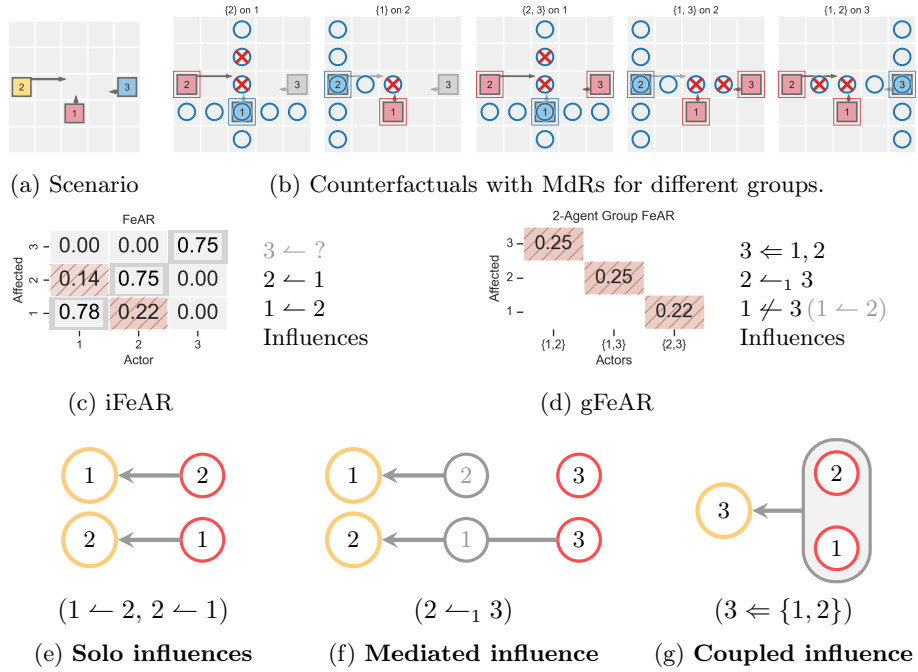
\begin{figure*}[tb]
    \centering
    \begin{subfigure}[t]{0.15\linewidth}
    \centering
        \includegraphics[width=\linewidth]{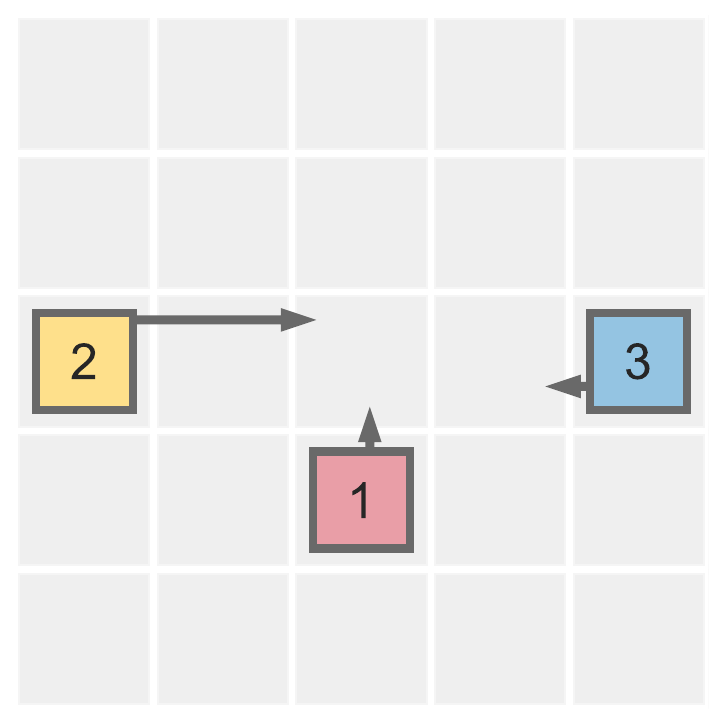}
        \caption{\footnotesize Scenario}
        \label{fig:SmallExampleScenario}
    \end{subfigure}
    \hfill
    \begin{subfigure}[t]{0.82\linewidth}
        \includegraphics[width=0.19\linewidth]{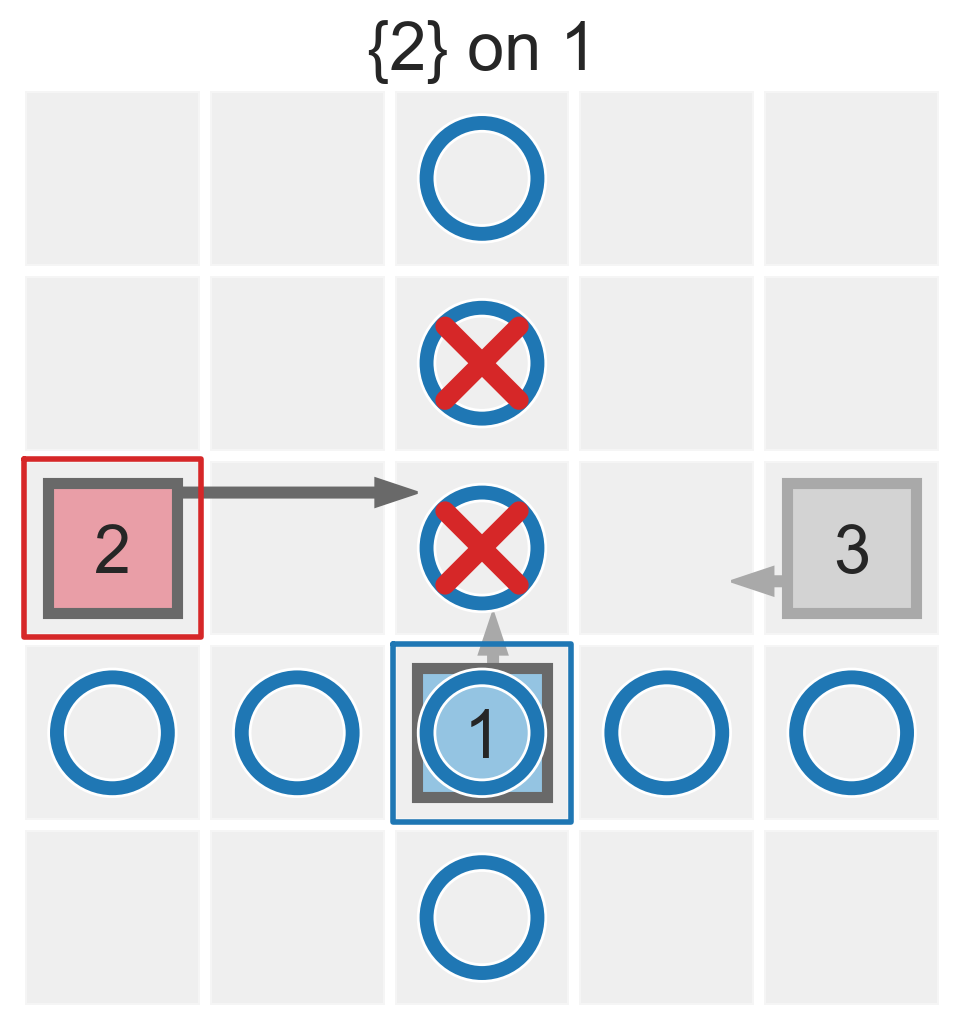}
        \includegraphics[width=0.19\linewidth]{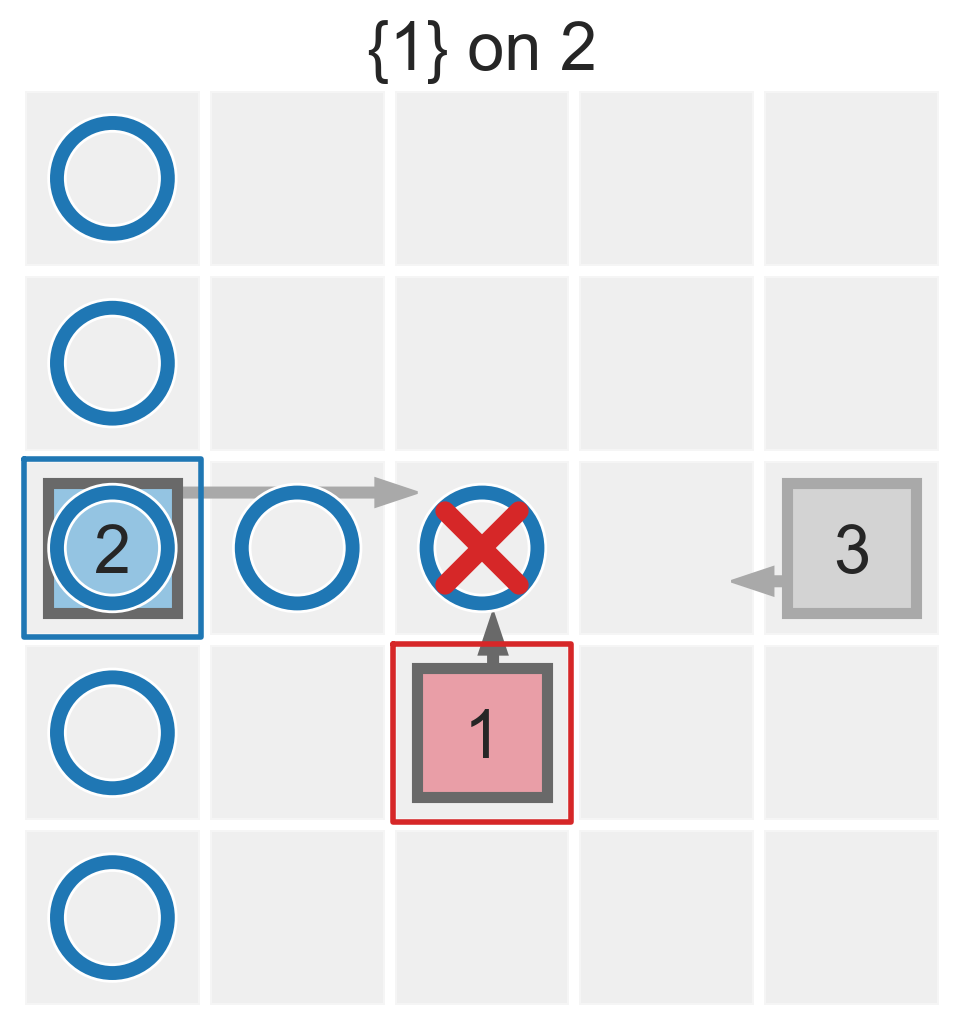}
        \includegraphics[width=0.19\linewidth]{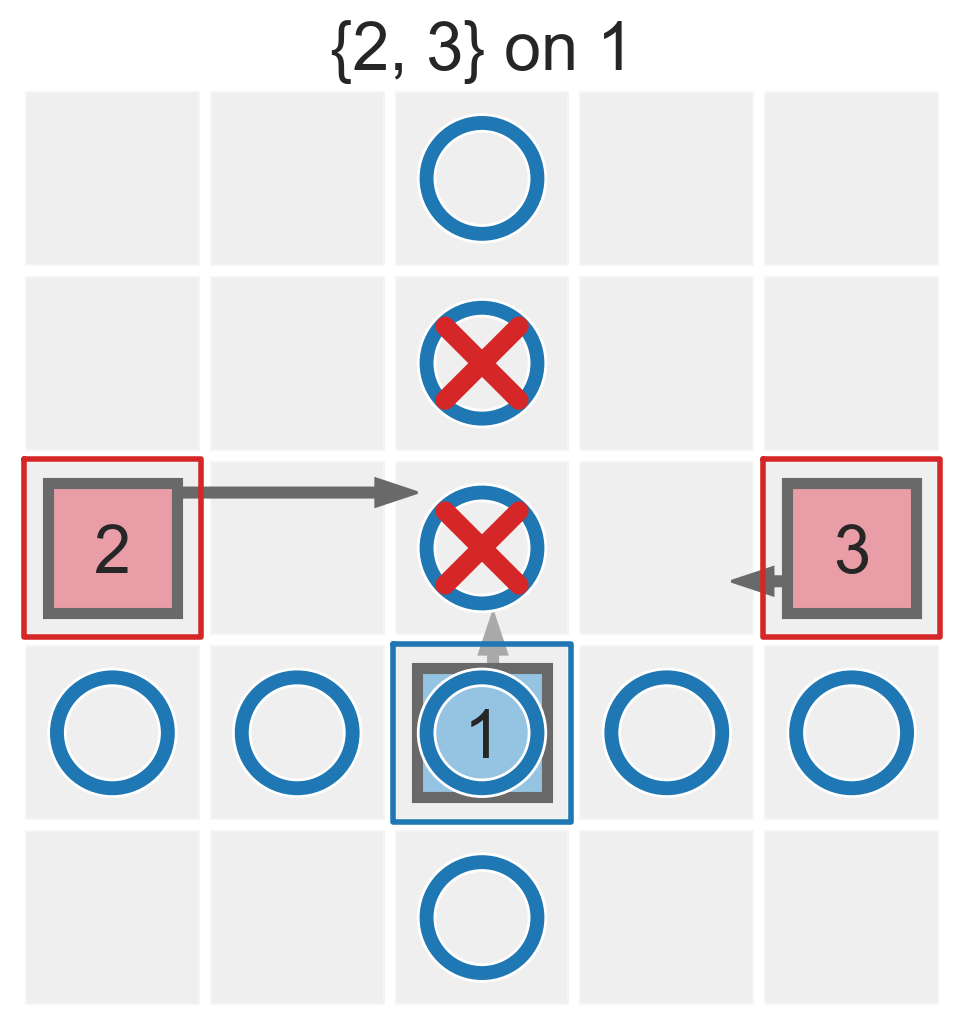}
        \includegraphics[width=0.19\linewidth]{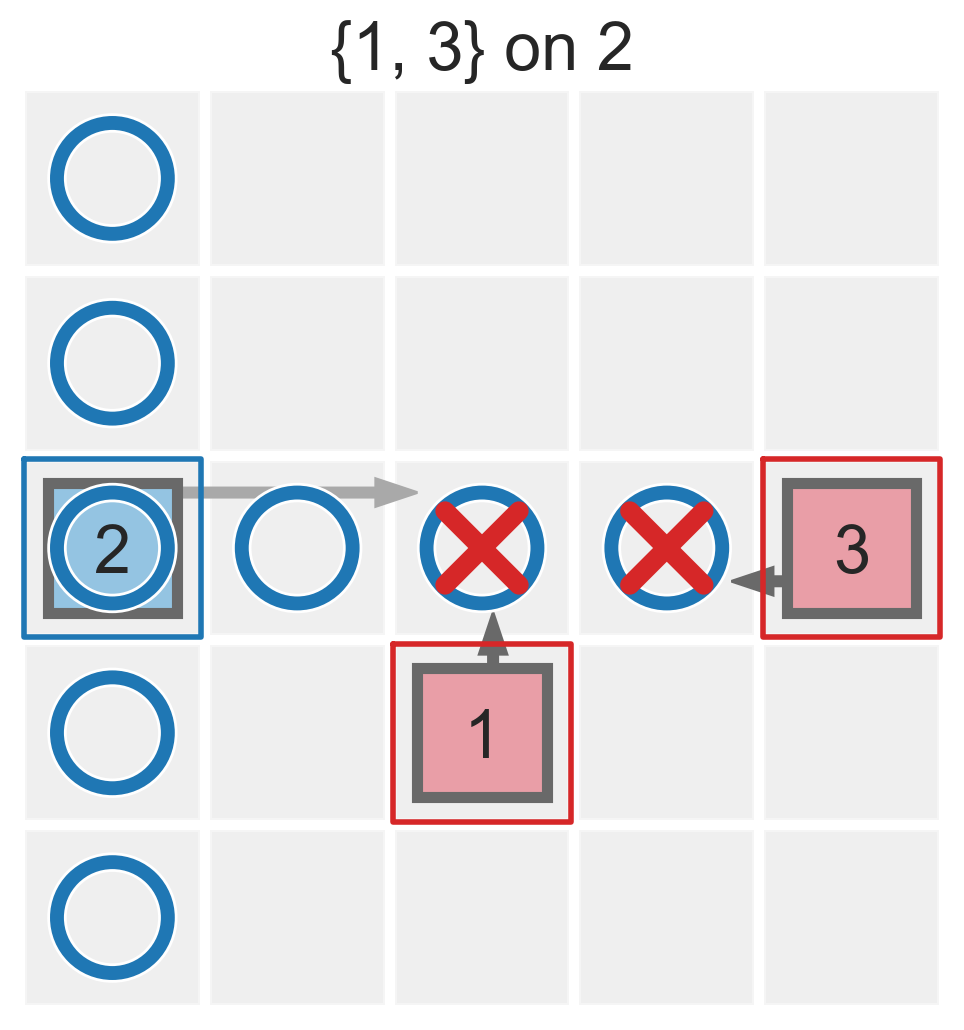}
        \includegraphics[width=0.19\linewidth]{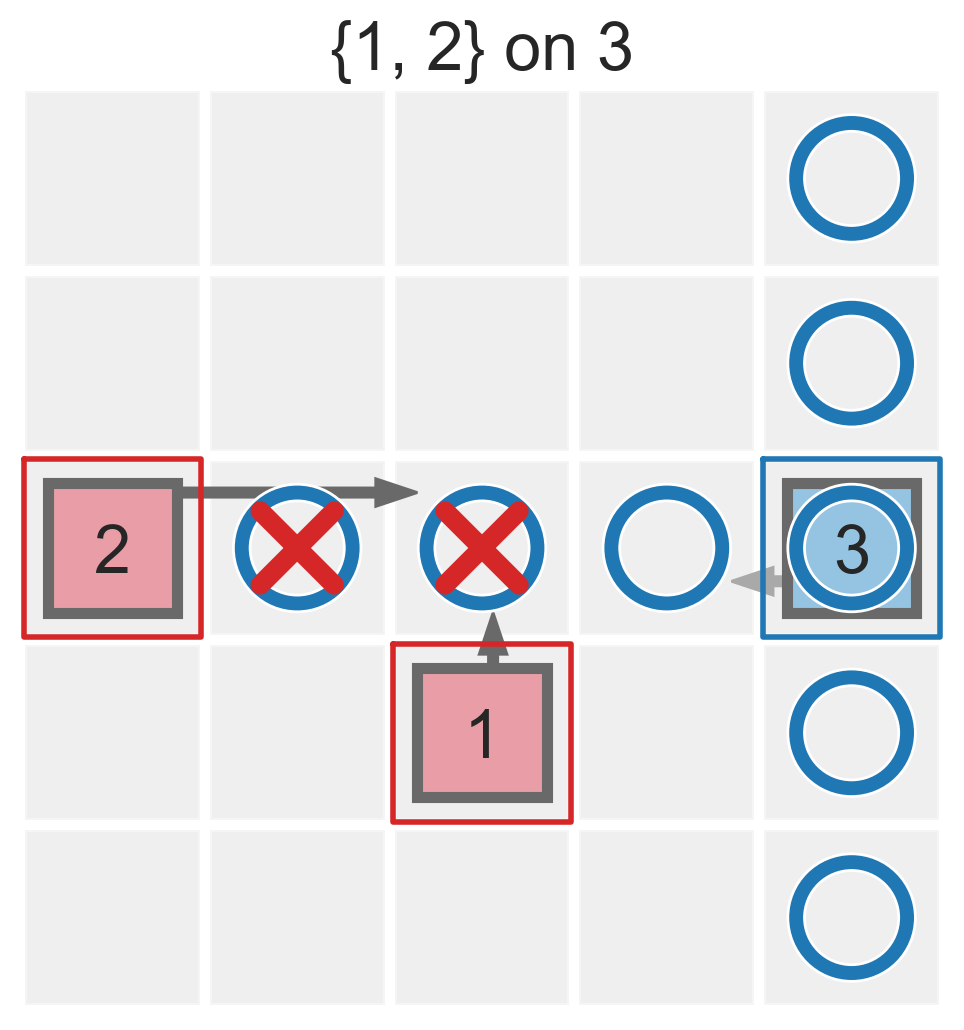}
        \caption{Counterfactuals with MdRs for different groups.}
        \label{fig:SmallExampleCounterfactuals}
    \end{subfigure}
    \begin{subfigure}[t]{0.25\linewidth}
    \centering
    \includegraphics[width=\linewidth]{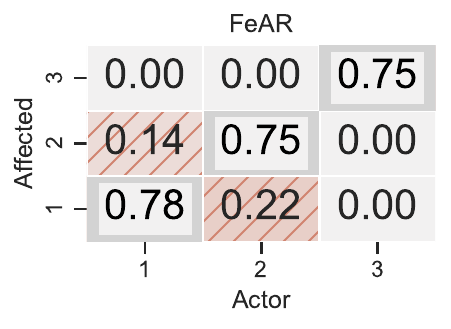}
    \caption{\footnotesize {iFeAR}}
    \label{fig:SmallExampleResults}
    \end{subfigure}
    \hfill
    \begin{subfigure}[b]{0.23\linewidth}
    \small
        $\textcolor{black!40}{\solo{~?}{3}}$\\
        $\solo{1}{2}$\\
        $\solo{2}{1}$\\
        ~ Influences\\
    \end{subfigure}
    \begin{subfigure}[t]{0.25\linewidth}
        \centering
        \includegraphics[width=\linewidth]{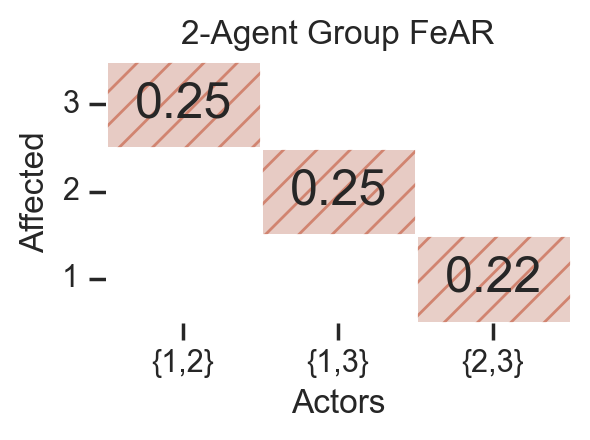}
        \caption{\footnotesize {gFeAR}}
        \label{fig:SmallExampleResultsGroup}
    \end{subfigure}
    \hfill
    \begin{subfigure}[b]{0.2\linewidth}
    \small
        $\coupled{1,2}{3}$\\
        $\mediated{3}{2}{1}$\\
        $1\not\leftharpoonup3 \, \textcolor{black!40}{(\solo{2}{1})}$ \\
        ~Influences\\
    \end{subfigure}
    \begin{subfigure}[b]{\linewidth}
    \begin{subfigure}[b]{0.3\linewidth}
    \centering
    % \raggedright
    \vspace{0.3cm}
        \footnotesize
\begin{tikzpicture}[
    affected/.style={circle, draw=Dandelion!70, fill=white, ultra thick, minimum size=0.7cm, font=\footnotesize},
    agent/.style={circle, draw=red!70, fill=white, very thick, minimum size=0.5cm, font=\footnotesize},
    group/.style={rectangle, rounded corners=12pt, draw=black!40, fill=black!5, very thick, inner sep=4pt},
    tier/.style={rectangle, draw=black!30, fill=black!10, inner sep=7pt},
    arrow/.style={->, >=stealth, ultra thick, draw=black!40, line cap=round},
    link/.style={- , ultra thick, draw=black!40, line cap=round}
]

    % Layer 1: Agent nodes (top)
    \node[affected] (AFFECTED) at (-1.5,0) {1};
    \node[agent] (T0G0A0) at (0,0.0) {2};
    % \node at (-5,0) {\normalsize$\solo{2}{1}$};

    \begin{pgfonlayer}{background}
        % Tier boundaries (bottom)
        % \node[tier, fit=(T0G0A0), label={[anchor=south]above:Tier 1}] (tier0) {};

    % Layer 2: Group boundaries (middle)
    \end{pgfonlayer}

    % Layer 4: Arrows (top)
    \draw[arrow] (T0G0A0) -- (AFFECTED);
\end{tikzpicture}
\normalsize
    \vspace{-1.2cm}
        \footnotesize
\begin{tikzpicture}[
    affected/.style={circle, draw=Dandelion!70, fill=white, ultra thick, minimum size=0.7cm, font=\footnotesize},
    agent/.style={circle, draw=red!70, fill=white, very thick, minimum size=0.5cm, font=\footnotesize},
    group/.style={rectangle, rounded corners=12pt, draw=black!40, fill=black!5, very thick, inner sep=4pt},
    tier/.style={rectangle, draw=black!30, fill=black!10, inner sep=7pt},
    arrow/.style={->, >=stealth, ultra thick, draw=black!40, line cap=round},
    link/.style={- , ultra thick, draw=black!40, line cap=round}
]

    % Layer 1: Agent nodes (top)
    \node[affected] (AFFECTED) at (-1.5,0) {2};
    \node[agent] (T0G0A0) at (0,0.0) {1};
    % \node at (-5,0) {\normalsize$\solo{2}{1}$};

    \begin{pgfonlayer}{background}
        % Tier boundaries (bottom)
        % \node[tier, fit=(T0G0A0), label={[anchor=south]above:Tier 1}] (tier0) {};

    % Layer 2: Group boundaries (middle)
    \end{pgfonlayer}

    % Layer 4: Arrows (top)
    \draw[arrow] (T0G0A0) -- (AFFECTED);
\end{tikzpicture}
\normalsize
    \vspace{-0.6cm}
    $\left(\solo{2}{1}, \, \solo{1}{2}\right)$
        \caption{\textbf{Solo influences}}
        \label{fig:solo_example}
    \end{subfigure}
    \begin{subfigure}[b]{0.38\linewidth}
    \centering
    \vspace{0.3cm}
        \footnotesize
\begin{tikzpicture}[
    affected/.style={circle, draw=Dandelion!70, fill=white, ultra thick, minimum size=0.7cm, font=\footnotesize},
    agent/.style={circle, draw=red!70, fill=white, very thick, minimum size=0.5cm, font=\footnotesize},
    mediator/.style={circle, draw=black!40, fill=white, very thick, minimum size=0.5cm, font=\footnotesize},
    group/.style={rectangle, rounded corners=12pt, draw=black!40, fill=black!5, very thick, inner sep=4pt},
    tier/.style={rectangle, draw=black!30, fill=black!10, inner sep=7pt},
    arrow/.style={->, >=stealth, ultra thick, draw=black!40, line cap=round},
    link/.style={- , ultra thick, draw=black!40, line cap=round}
]

    % Layer 1: Agent nodes (top)
    \node[affected] (AFFECTED) at (-1.5,0) {1};
    \node[mediator] (T0G0A0) at (0,0.0) {\textcolor{black!40}{2}};
    \node[agent] (T1G0A0) at (1.5,0.0) {3};
    % \node at (-5,0) {\normalsize$\mediated{2}{3}{1}$};

    \begin{pgfonlayer}{background}
        % % Tier boundaries (bottom)
        % \node[tier, fit=(T0G0A0), label={[anchor=south]above:Tier 1}] (tier0) {};
        % \node[tier, fit=(T1G0A0), label={[anchor=south]above:Tier 2}] (tier1) {};

    % Layer 2: Group boundaries (middle)
    \end{pgfonlayer}

    % Layer 4: Arrows (top)
    % \draw[link] (T1G0A0) -- (T0G0A0);
    \draw[arrow] (T0G0A0) -> (AFFECTED);
\end{tikzpicture}
\normalsize
    \vspace{-1.2cm}
        \footnotesize
\begin{tikzpicture}[
    affected/.style={circle, draw=Dandelion!70, fill=white, ultra thick, minimum size=0.7cm, font=\footnotesize},
    agent/.style={circle, draw=red!70, fill=white, very thick, minimum size=0.5cm, font=\footnotesize},
    mediator/.style={circle, draw=black!40, fill=white, very thick, minimum size=0.5cm, font=\footnotesize},
    group/.style={rectangle, rounded corners=12pt, draw=black!40, fill=black!5, very thick, inner sep=4pt},
    tier/.style={rectangle, draw=black!30, fill=black!10, inner sep=7pt},
    arrow/.style={->, >=stealth, ultra thick, draw=black!40, line cap=round},
    link/.style={- , ultra thick, draw=black!40, line cap=round}
]

    % Layer 1: Agent nodes (top)
    \node[affected] (AFFECTED) at (-1.5,0) {2};
    \node[mediator] (T0G0A0) at (0,0.0) {\textcolor{black!40}{1}};
    \node[agent] (T1G0A0) at (1.5,0.0) {3};
    % \node at (-5,0) {\normalsize$\mediated{2}{3}{1}$};

    \begin{pgfonlayer}{background}
        % % Tier boundaries (bottom)
        % \node[tier, fit=(T0G0A0), label={[anchor=south]above:Tier 1}] (tier0) {};
        % \node[tier, fit=(T1G0A0), label={[anchor=south]above:Tier 2}] (tier1) {};

    % Layer 2: Group boundaries (middle)
    \end{pgfonlayer}

    % Layer 4: Arrows (top)
    \draw[link] (T1G0A0) -- (T0G0A0);
    \draw[arrow] (T0G0A0) -> (AFFECTED);
\end{tikzpicture}
\normalsize
    \vspace{-0.6cm}
    $\left(\mediated{3}{2}{1}\right)$
        \caption{\textbf{Mediated influence}}
        \label{fig:mediated_example}  
    \end{subfigure}
    \begin{subfigure}[b]{0.3\linewidth}
    \centering
        \footnotesize
\begin{tikzpicture}[
    affected/.style={circle, draw=Dandelion!70, fill=white, ultra thick, minimum size=0.7cm, font=\footnotesize},
    agent/.style={circle, draw=red!70, fill=white, very thick, minimum size=0.5cm, font=\footnotesize},
    group/.style={rectangle, rounded corners=12pt, draw=black!40, fill=black!5, very thick, inner sep=4pt},
    tier/.style={rectangle, draw=black!30, fill=black!10, inner sep=7pt},
    arrow/.style={->, >=stealth, ultra thick, draw=black!40, line cap=round},
    link/.style={- , ultra thick, draw=black!40, line cap=round}
]

    % Layer 1: Agent nodes (top)
    \node[affected] (AFFECTED) at (-1.5,0) {3};
    \node[agent] (T0G0A0) at (0,0.5) {2};
    \node[agent] (T0G0A1) at (0,-0.5) {1};
    % \node at (-5,0) {\normalsize$\coupled{\{1,2\}}{3}$};

    \begin{pgfonlayer}{background}
        % Tier boundaries (bottom)
        % \node[tier, fit=(T0G0A0)(T0G0A1), label={[anchor=south]above:Tier 1}] (tier0) {};

    % Layer 2: Group boundaries (middle)
        \node[group, fit=(T0G0A0)(T0G0A1)] (group0_0) {};
    \end{pgfonlayer}

    % Layer 4: Arrows (top)
    \draw[arrow] (group0_0) -- (AFFECTED);
\end{tikzpicture}
\normalsize
        \vspace{-0.8cm}
        $\left(\coupled{\{1,2\}}{3}\right)$
        \caption{\textbf{Coupled influence}}
        \label{fig:coupled_example}
    \end{subfigure}
    \end{subfigure}
    \normalsize
\caption{
\small
\textbf{Types of assertive influence based on group FeAR:} 
For the illustrative scenario~(a) where three agents are moving towards each other, we show how counterfactuals based on the MdR of actors~(b) are used to compute iFeAR values~(c) for individual actors and gFeAR values for group actors~(d). While iFeAR can only identify \emph{solo} influences  ($\solo{2}{1}$ and $\solo{1}{2}$) (e), analysing gFeAR can reveal additional \emph{mediated} ($\mediated{3}{2}{1}$)~(f) and \emph{coupled} ($\coupled{\group{1,2}}{3}$)~(g) influences. Even though $\FeAR_{3,3}<1$ reflects the reduction in feasible action space of 3, iFeAR cannot identify the assertive actors; which are revealed by gFeAR ($\coupled{\group{1,2}}{3}$). Also note that agent 3 has no influence on agent 1 as $\FeAR_{2,1}=\FeAR_{\group{2,3},1}$.
}
\label{fig:SmallExample}
\end{figure*}

The function $Z(x)$ clips the values of FeAR to $\left[-1,1\right]$ to aid interpretability of $\FeAR$ values and $\epsilon$ in the denominator ensures that FeAR is defined when $\nijMdR=0$.
Positive values of $\FeAR_{i,j}$ indicate that actor $i$ is being assertive towards the affected agent $j$ and is hence causally responsible for $j$'s trajectory. For example, in the scenario shown in \cref{fig:SmallExample}, agent 1 decreases the feasible action space of agent 2 by one (\cref{fig:SmallExampleCounterfactuals}) and hence $\FeAR_{1,2}>0$ (\cref{fig:SmallExampleResults}).

\subsection{Group FeAR}
\label{sec:GroupFeAR}

To better capture causal responsibility in cases of causal overdeterminism, we propose the FeAR metric for groups (gFeAR) as follows:

\begin{definition}[\textbf{Group FeAR}]
    The Feasible Action-Space Reduction $(\FeAR)$ imposed by a non-empty group $G \subseteq \neg j$ on an affected agent $j$ , is defined as:
\end{definition}
\begin{equation}
        \FeAR_{G, j}(\aState,\jointAction) = \frac{\nijMdRG - \nij}{\nijMdRG + \epsilon}.
\end{equation}

Looking at \cref{fig:SmallExample} again, the group $\{1,3\}$ acting together reduces the feasible action space of 2 by 2 \cref{fig:SmallExampleCounterfactuals}. This is captured by the definition of FeAR for groups and  $\FeAR_{\group{1,3},2}>\FeAR_{1,2}>0$ shows that the group of actors $\group{1,3}$ has more influence on agent 2 than just the individual actor 1. Furthermore, when considering the effect on agent 3, $\FeAR_{1,3}=\FeAR_{2,3}=0$, fails to capture any influence of other agents. But $\FeAR_{\group{1,2},3}>0$ shows that collectively $\group{1,2}$ is behaving assertively towards agent 3 and is hence collectively causally responsible for the trajectory of agent 3. Thus, gFeAR can quantify casual responsibility in cases of  causal overdetermination.

As seen above, agents can have different influences on an affected agent when considered to be acting individually or as part of a group. The following section categorises these types of influences.

\subsection{Types of assertive influence}
\label{sec:TypesOfInfluence}

Based on whether agents have assertive influence on their own or as part of groups, we have identified 3 fundamental types of assertive influence: \emph{solo influence}, \emph{mediated influence} and \emph{coupled influence}, and a fourth derived type of influence \emph{mediated coupled influence}.
These four types of influence defined below span the whole spectrum of assertive influences and helps us compare the assertiveness of different agents.

\begin{definition}[Solo influence $\solo{i}{j}$]

Agent $i$ has solo influence on agent $j$ if $\FeAR_{i,j} > 0$.
\end{definition}

\begin{definition}[\textbf{Mediated influence} $\mediated{i}{j}{G}$]
Agent $i$ has a mediated influence on agent $j$ if 
$\FeAR_{i,j} = 0$, and $\exists G \subset \neg j \setminus \singleton{i}$, $G \neq \emptyset $ such that 
$\FeAR_{G \cup \singleton{i} ,j} > \FeAR_{G,j} >0$.
\end{definition}

\begin{definition}[\textbf{Coupled influence} $\coupled{G}{j}$]
    
All agents in group $G \subset \neg j \setminus \singleton{i}$, $G \neq \emptyset $ have coupled influence on agent $j$ if $\FeAR_{i,j} = 0 \quad \forall i \in G$ and $\FeAR_{G,j} > 0$. 
\end{definition}

There can also be cases where a group of agents have coupled influence that is mediated by another group of agents. Therefore a more general definition of mediated coupled influence is as follows:

\begin{definition}[\textbf{Mediated coupled influence} $\mediatedCoupled{G}{j}{G'}$]

Group $G$ has a coupled influence on agent $j$ which is mediated by another group $G'$
if $\FeAR_{i,j} = \FeAR_{G',j} \quad \forall i \in G$ and $\FeAR_{G' \cup G,j} > \FeAR_{G',j}$ 
\end{definition}

In the example shown in \cref{fig:SmallExample}, agents 1 and 2 have solo influences on each other ($\solo{2}{1}$, $\solo{1}{2
}$),
agent 3 has an influence on agent 2 which is mediated by agent 1 ($\mediated{3}{2}{1}$),
and
agents 1 and 2 have a coupled influence on agent 3 ($\coupled{\group{1,2}}{3}$). \cref{fig:solo_example,fig:mediated_example,fig:coupled_example} show how these influences are pictorially represented. 

\cref{fig:tiers-illustration} shows a more complicated scenario with more intricate influences. For example, the group $\group{5,7}$ has a mediated coupled influence on agent 1 $\mediatedCoupled{\group{5,7}}{1}{\group{2,3,4,6}}$ (\cref{fig:tiers-illustration-tiers}). For systematically unravelling these intricate dependencies of influence by identifying minimal groups that are causally responsible, we propose a tiering algorithm for ranking the assertive influences on each affected agent.

\begin{algorithm}
\caption{Tiering the assertive influence of agents}
\label{alg:tiers}
\scriptsize
% \footnotesize
% \vspace{-0.2cm}
\[\begin{array}{l r}
\courteous{j} \leftarrow \{i:\FeAR_{i,j}<0\} & \text{Find courteous agents $\courteous{j}$.}\\
\candidate{j} \leftarrow \neg j \setminus\courteous{j} & \text{Remove $\courteous{j}$ from candidate agents $\candidate{j}$.}\\
\mathcal{R}_0 \leftarrow \emptyset & \text{$\mathcal{R}_n$ --- Assertive agents until $\tier{j}{n}$.}\\

\textbf{for } n=1,2,\dots: & \text{For each tier,}\\
\quad \tier{j}{n} \leftarrow \emptyset & \\
\quad \textbf{for } k=1,\dots,|\candidate{j}|: & \text{starting with k=1,}\\
\quad\quad \textbf{for all } G\subseteq\candidate{j}, |G|=k: & \text{loop through groups with k agents,}\\
\quad\quad\quad \textbf{if } \FeAR_{\mathcal{R}_{n-1}\cup G,j}>\FeAR_{\mathcal{R}_{n-1},j}: &\text{identify assertive groups,} \\
\quad\quad\quad\quad \tier{j}{n} \leftarrow \tier{j}{n}\cup\{G\},\; & \text{add them to tiers, and}\\
\quad\quad\quad\quad \candidate{j} \leftarrow \candidate{j}\setminus G & \text{remove them from candidate agents.}\\
\quad \textbf{if } \tier{j}{n}=\emptyset: \textbf{ break} & \\
\quad \mathcal{R}_n \leftarrow \mathcal{R}_{n-1}\cup\bigcup_{G\in\tier{j}{n}} G &
\end{array}
\]
\vspace{-0.2cm}
\normalsize
\end{algorithm}

\subsection{Ranking agents based on influence}
\label{sec:RankingAgents}

In cases of agents having mediated influence, their influence is conditional to the actions of the mediating agent. Thus, based on the intuition that the assertive influence of agents with mediated influence should be ranked lower than the assertive influence of those agents that mediate these mediated influences, we propose a tiering algorithm for ranking the influence of different agents into tiers (\cref{alg:tiers}) \footnote{Detailed implementation of the algorithm can be found at \url{github.com/DAI-Lab-HERALD/FeAR}.}.

To represent how higher tiers have more influence on the affected agent $j$, we use the $\succ$ operator :
$\tier{j}{n} \succ \tier{j}{n+1}$.

\begin{figure*}[tb]
    \centering
    \begin{subfigure}[c]{0.22\linewidth}
    \centering
        \begin{subfigure}[c]{\linewidth}
        \centering
        \includegraphics[width=0.8\linewidth]{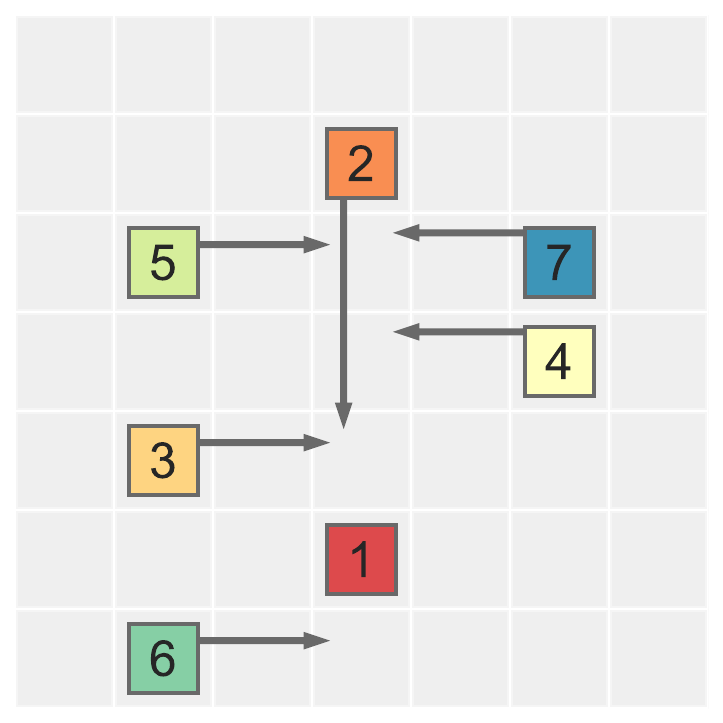}
        \caption{Scenario}
        \label{fig:tiers-illustration-scenario}
        \end{subfigure}
        \begin{subfigure}[c]{\linewidth}
        \includegraphics[width=\linewidth]{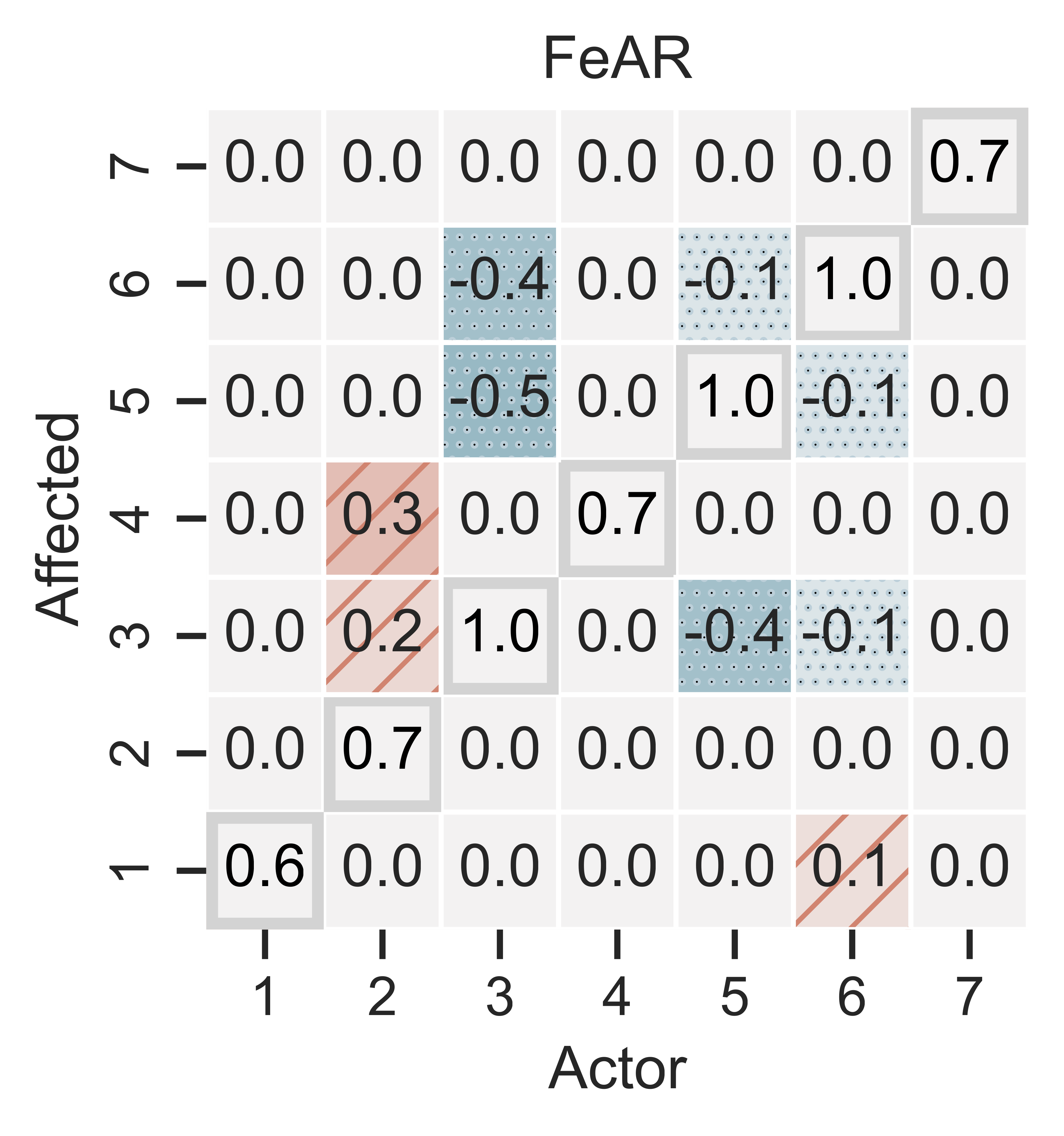}
        \caption{iFeAR}
        \label{fig:tiers-illustration-fear}
        \end{subfigure}
    \end{subfigure}
    \hspace{0.1cm}
    \begin{subfigure}[c]{0.4\linewidth}
        \centering
        \begin{subfigure}[c]{0.48\linewidth}
        \includegraphics[width=\linewidth]{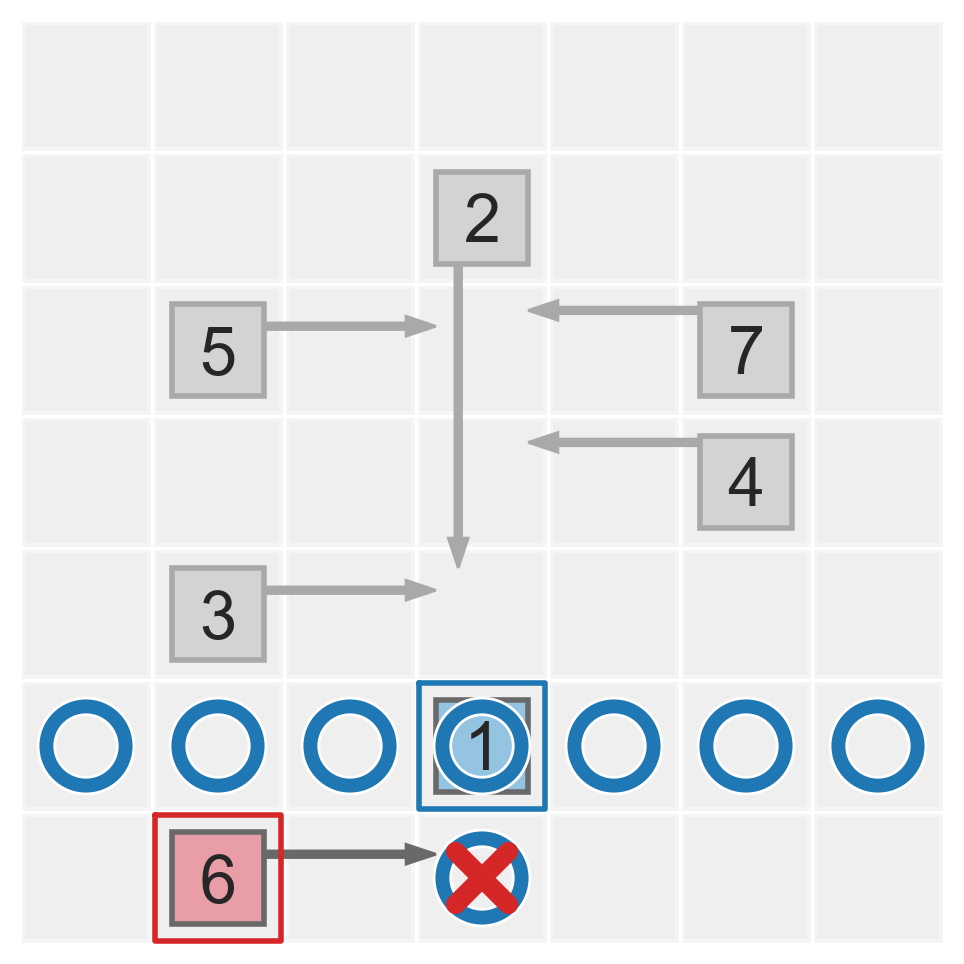}
        \end{subfigure}
        \begin{subfigure}[c]{0.48\linewidth}
        \includegraphics[width=\linewidth]{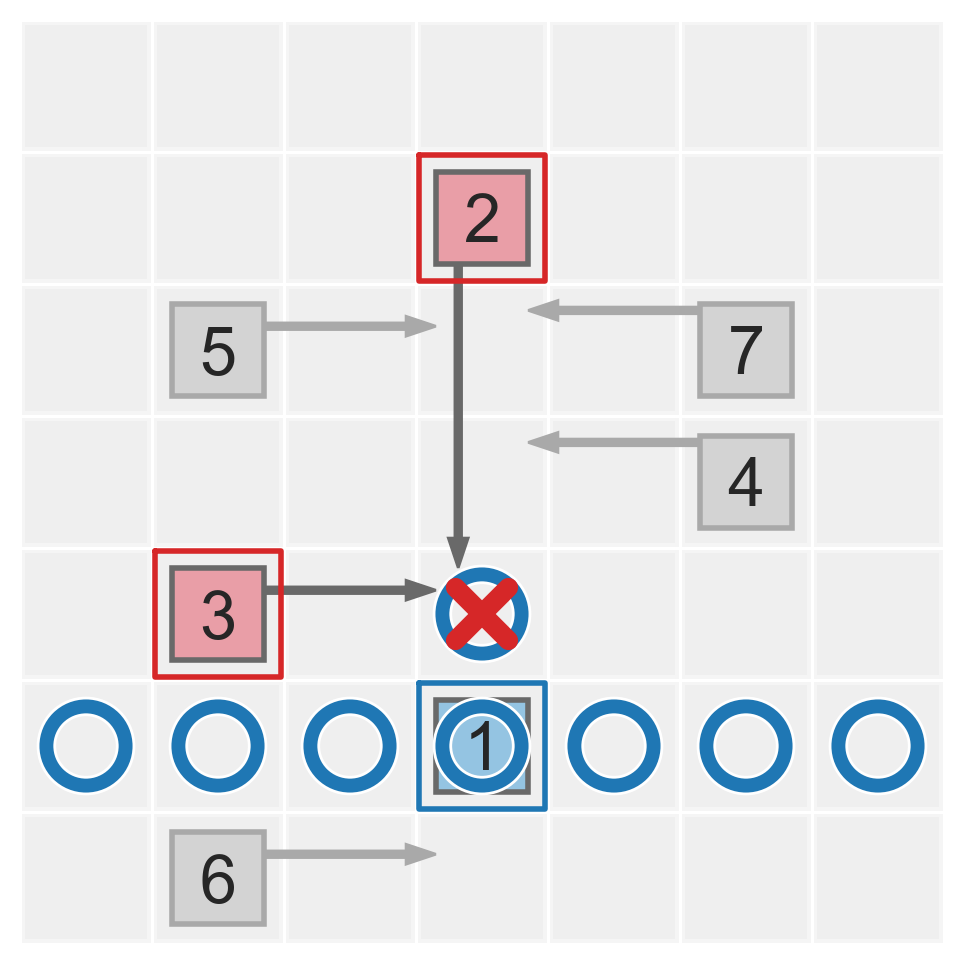}
        \end{subfigure}
        \begin{subfigure}[c]{0.48\linewidth}
        \includegraphics[width=\linewidth]{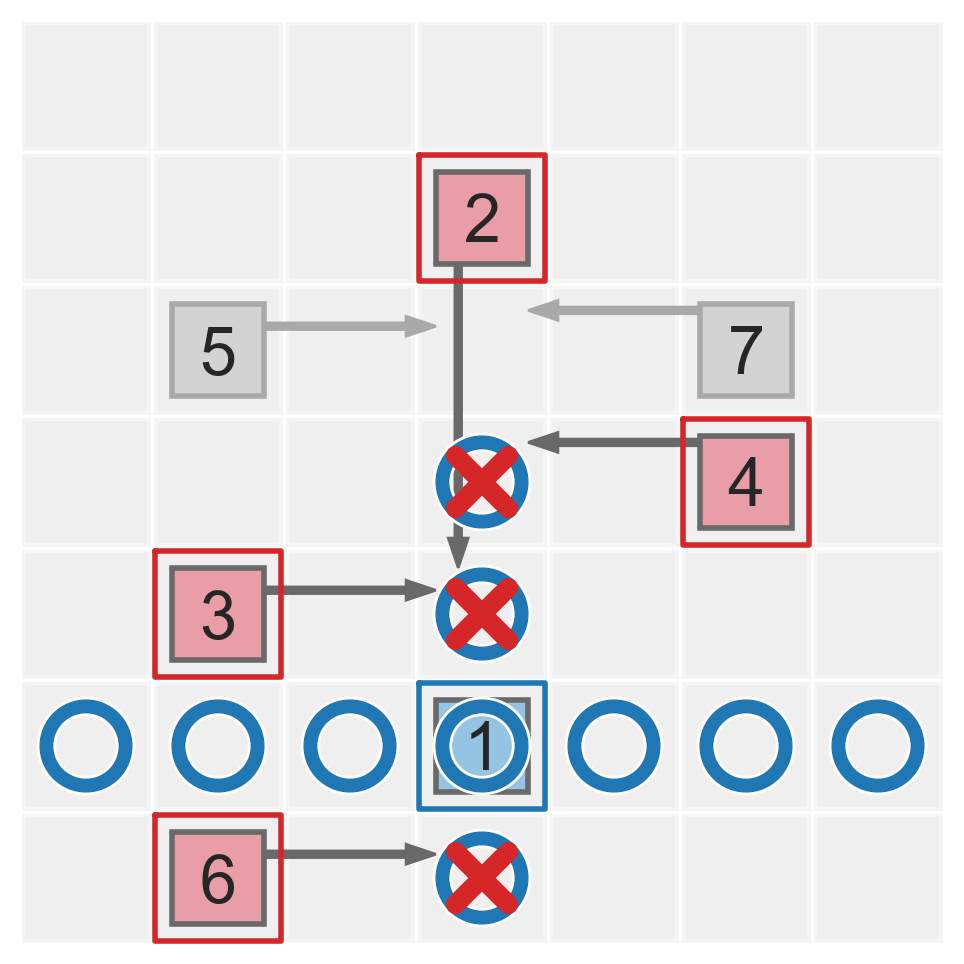}
        \end{subfigure}
        \begin{subfigure}[c]{0.48\linewidth}
        \includegraphics[width=\linewidth]{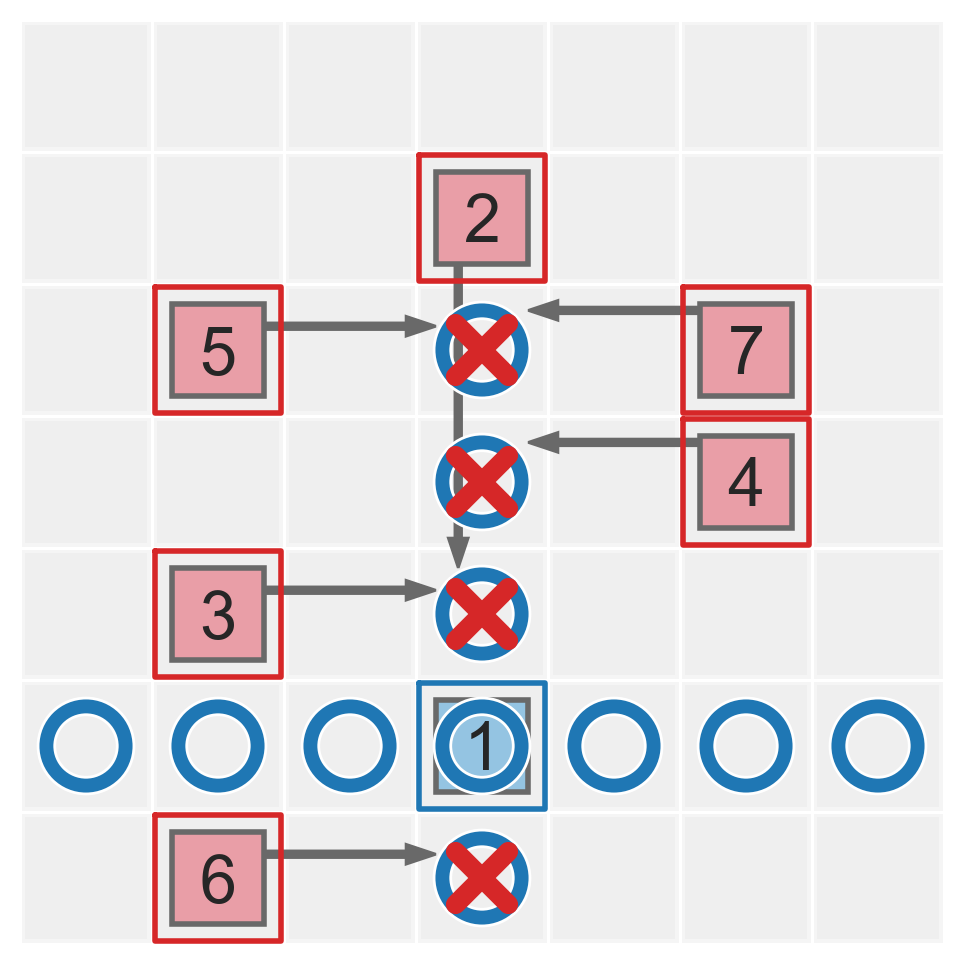}
        \end{subfigure}
        \caption{Assertive influences on agent 1.}
        \label{fig:tiers-illustration-effect-on-1}
    \end{subfigure}
    \begin{subfigure}[c]{0.3\linewidth}
        \centering
        \footnotesize
\begin{tikzpicture}[
    affected/.style={circle, draw=Dandelion!70, fill=white, ultra thick, minimum size=0.6cm, font=\scriptsize},
    agent/.style={circle, draw=red!70, fill=white, very thick, minimum size=0.4cm, font=\scriptsize},
    group/.style={rectangle, rounded corners=10pt, draw=black!40, fill=black!5, very thick, inner sep=3pt},
    tier/.style={rectangle, draw=black!30, fill=black!10, inner sep=5pt},
    arrow/.style={->, >=stealth, ultra thick, draw=black!40, line cap=round},
    link/.style={-, >=stealth, ultra thick, draw=black!40, line cap=round},
]

    % Layer 1: Agent nodes (top)
    \node[affected] (AFFECTED) at (-1.1,0) {1};
    \node[agent] (T0G0A0) at (0,0.8) {6};
    \node[agent] (T0G1A0) at (0,-0.0) {2};
    \node[agent] (T0G1A1) at (0,-0.8) {3};
    \node[agent] (T1G0A0) at (1.1,0.0) {4};
    \node[agent] (T2G0A0) at (2.2,0.4) {5};
    \node[agent] (T2G0A1) at (2.2,-0.4) {7};

    \begin{pgfonlayer}{background}
        % Tier boundaries (bottom)
        \node[tier, fit=(T0G0A0)(T0G1A0)(T0G1A1), label={[anchor=south]above:$\tier{1}{1}$ }] (tier0) {};
        \node[tier, fit=(T1G0A0), label={[anchor=south]above:$\tier{1}{2}$ }] (tier1) {};
        \node[tier, fit=(T2G0A0)(T2G0A1), label={[anchor=south]above:$\tier{1}{3}$ }] (tier2) {};

    % Layer 2: Group boundaries (middle)
        \node[group, fit=(T0G1A0)(T0G1A1)] (group0_1) {};
        \node[group, fit=(T2G0A0)(T2G0A1)] (group2_0) {};
    \end{pgfonlayer}

    % Layer 4: Arrows (top)
    \draw[link] (group2_0) -- (tier1);
    \draw[link] (T1G0A0) -- (tier0);
    \draw[arrow] (T0G0A0) -- (AFFECTED);
    \draw[arrow] (group0_1) -- (AFFECTED);
\end{tikzpicture}
\normalsize
        \caption{Tiers affecting agent 1}
        \label{fig:tiers-illustration-tiers}
        \end{subfigure}
        \hfill
    \caption{
    \small
    \textbf{Ranking the assertiveness of agents into tiers $\tier{j}{n}$:} 
    In this illustrative scenario~(a), when considering agent 1 as the affected, iFeAR only show agent 6 as being the assertive (b). However, counterfactuals with groups (c) reveal more assertive influences on agent 1 which are systematically ranked into tiers  $\tier{1}{n}$~(d).
    \vspace{-0.5cm}
    }
    \label{fig:tiers-illustration}
\end{figure*}

Consider how different actors influence agent 1  for the example in \cref{fig:tiers-illustration-scenario}. 
\cref{fig:tiers-illustration-effect-on-1} 
shows the assertive influence of groups of actors $\group{6}$, $\group{2,3}$, $\group{2,3,4,6}$ and $\group{2,3,4,5,6,7}$. 
Based on these assertive influences, the algorithm systematically identifies 
the solo influence of agent 6 ($\solo{6}{1}$), the coupled influence of group $\group{2,3}$ ($\coupled{\group{2,3}}{1}$), the mediated influence of agent 4 ($\mediated{4}{1}{\group{2,3,6}}$) and the mediated coupled influence of group $\group{5,7}$ ($\mediatedCoupled{\group{5,7}}{1}{\group{2,3,4,6}}$), and sorts these assertive influences into three tiers $6\sim2\sim3\succ4\succ5\sim7$ as shown in \cref{fig:tiers-illustration-tiers}. 
Thus, the tiers provide richer information about assertive influences than the assertive influences found from the positive values of individual FeAR (\cref{fig:tiers-illustration-fear}).  

\section{Scenario-based simulations}
\label{sec:ScenarioSims}
% We used scenario-based simulations to demonstrate the benefits of considering the assertive influences of groups.
We defined metrics to quantify how group FeAR can uncover more information about an interaction than individual FeAR~(\cref{sec:Metrics}). Using these metrics, we explored group effects in simulations of different scenarios~(\cref{sec:Scenarios}).

\subsection{Metrics}
\label{sec:Metrics}
Since the goal of group FeAR and the tiering algorithm is to indentify assertive agents and to rank their assertiveness,
we consider two metrics for comparing FeAR and group FeAR, 1) based on the number of assertive agents and 2) based on the alignment of rankings of assertiveness using Kendall's $\tau$. To better understand the relationship of these metrics to the proximity of agents, we plot these against the median Manhattan distance to the affected agent.

\noindent\textbf{Difference in the number of assertive agents:}
For an affected agent $j$, the number of assertive agents identified with individual FeAR  is the number of actors $i$ with $\FeAR_{i,j}>0$, and the number of assertive agents identified group FeAR is the number of actors in tiers $\tier{j}{n}$.
Since, the number of assertive agents vary with scenarios, we use the difference in the number of assertive agents identified using group FeAR and individual FeAR to provide a consistent metric across scenarios:
\begin{equation}
\deltaAssertive = 
\nAssertivegFeAR - \nAssertiveiFeAR
=
\left| \cup_n \tier{j}{n} \right| - 
\left|\left\{ i: \FeAR_{i,j}>0 \right\}\right|.
\end{equation}

\noindent\textbf{Kendall's tau for rankings:}
For each affected agent, we can rank the assertiveness of other agents in three ways: 1) $\fearRanks$ ranks: by sorting the positive values of individual FeAR, 2) $\tierRanks$ ranks: based on the tiers generated by the tiering algorithm based on group FeAR, and 3) $\shapRanks$ ranks: based on ranking positive Shapely values~\cite{shapleyValueNPersonGames1953} generated from all the group FeAR values for that affected agent. 

We compare $\fearRanks$ and $\tierRanks$ ranks against baseline of $\shapRanks$ derived from Shapely values which are the state of the art when computing individual contributions to groups \cite{dobzinskiShapleyCostSharing2018,shapleyValueNPersonGames1953,yazdanpanahStrategicResponsibilityImperfect2019}.  
Shapley values are calculated based on the marginal changes in group FeAR value when an actor is added to a group of actors~\cite{shapleyValueNPersonGames1953}. 
The main difference between $\shapRanks$ ranks and $\tierRanks$ ranks is that shapely values consider all possible ways for assembling groups, while tiers are constructed in a more systematic manner starting from solo and coupled influences and then moving onto mediated influences.
% We include $\shapRanks$ as a baseline to compare the $\tierRanks$ ranks generated by our algorithm for ranking assertiveness.

For each affected agent in each case, we compare two rankings of assertiveness using Kendall's tau~\cite{kendallNewMeasureRank1938} which returns $\tau=+1$ if all pairs of actors have the same relative ranks in the both rankings. When creating the arrays of ranks for comparisons, non-assertive agents are given a rank of $k+1$ where $k$ is the total number of agents in that scenario.

The rankings of assertive influences, as in the case of the number of assertive agents, are dependent on the scenarios. To better understand the relationship between specific scenarios and these metrics we plot them against a metric for the proximity of agents.

\noindent\textbf{Median Manhattan Distance:}
All the scenarios for the randomised simulations~(\cref{sec:Scenarios}) had the same number of agents, and we used the median Manhattan distance between agents to indicate the proximity of agents in each case.
Cases where agents are closer together would have lower median Manhattan distances.
By plotting the differences in number of assertive agents identified and Kendall's $\tau$ comparing ranks for different median Manhattan distances, we can see how group effects are related to proximity of agents.

\subsection{Scenarios}
\label{sec:Scenarios}

We start by analysing a particular scenario (S1) in detail to illustrate how the metrics capture group effects. Later, we use these metrics to explore group effects in pseudo-randomised simulations of three different scenarios (S2, S3, S4).

\noindent\textbf{S1:Robot crossing pedestrians:}
For the detailed scenario, we consider the description in the introduction where a robot (agent 5) is crossing the path of some pedestrians \cref{fig:CrossyRoadIntro} and crashing into pedestrians 2, 3, 4 and 7. 

\noindent\textbf{S2-4:Randomised simulations:}
We consider three scenarios \emph{S2:Aggressive}, \emph{S3:Directed} and \emph{S4:Random} for the pseudo-randomised simulations (see \cref{fig:sims-scenarios}).
For \emph{Aggressive}, agents spawn in random locations and the policy makes agents take random actions to aggressively cross the intersection. For \emph{Directed}, agents spawn in fixed locations to the left of the intersection and then the policy makes them randomly choose actions to gently cross to the right and slow down after crossing the intersection. For \emph{Random}, agents spawn in the same location as \emph{Directed}, but then take completely random actions in all directions. 50 simulations of each scenario were run with 5 iterations per simulation. Thus, in total there were 250 separate cases for each scenario with eight agents.

\begin{figure}[tb]
    \centering
    \begin{subfigure}[c]{\linewidth}
        \begin{subfigure}[c]{0.38\linewidth}
        \centering
        \includegraphics[width=\linewidth]{Figures/Plots/CrossyRoad1_scenario.pdf}
        \caption{\footnotesize Robot (5) crossing pedestrians (1, 2, 3, 4, 6, 7 and 8)}
        \end{subfigure}
        \hfill
        \begin{subfigure}[c]{0.3\linewidth}
        \raggedleft
        \includegraphics[width=0.9\linewidth]{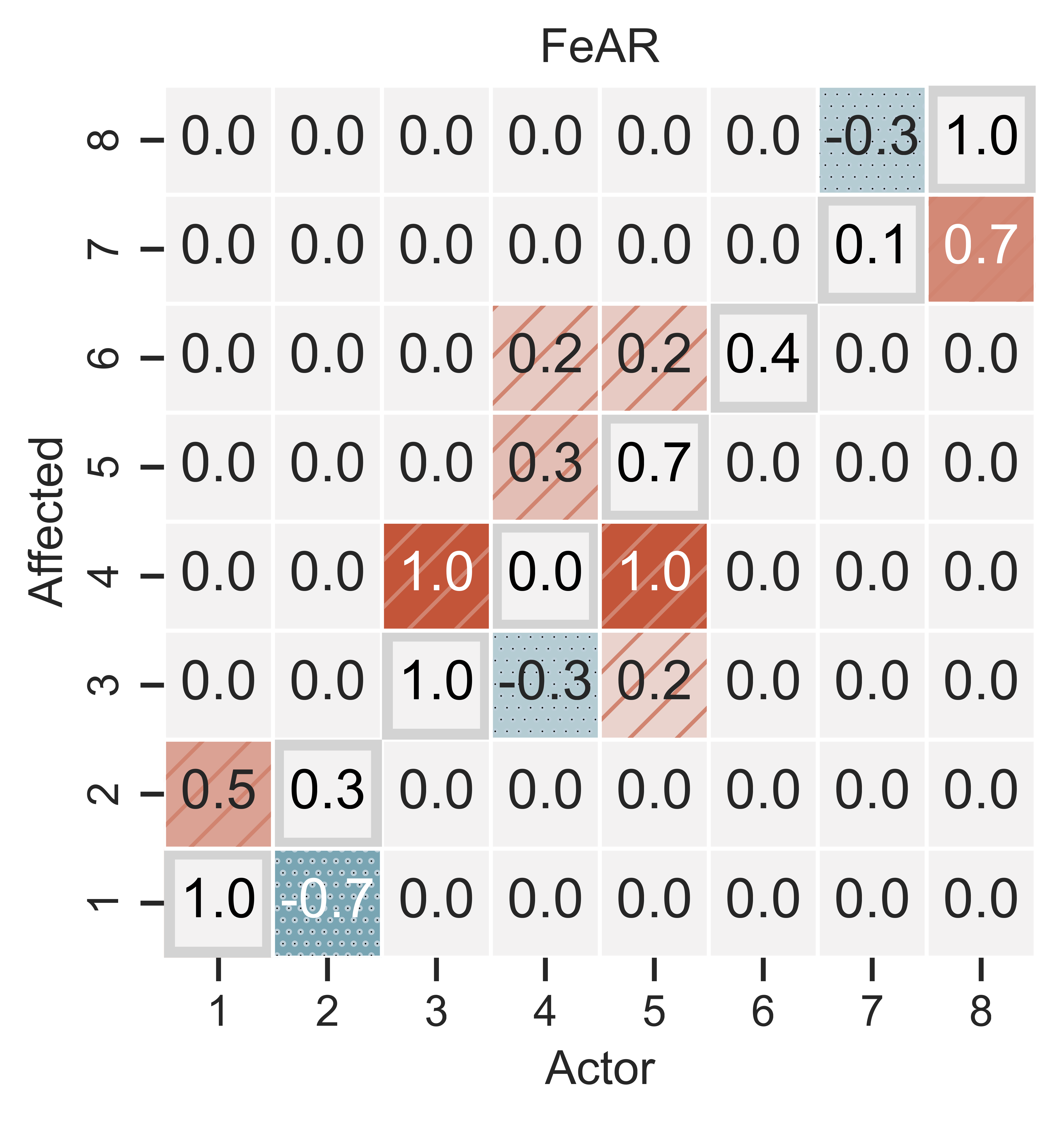}
        \vspace{-0.15cm}
        \caption{\footnotesize iFeAR}
        \label{fig:CrossyRoad-FeAR}
        \end{subfigure}
        \begin{subfigure}[c]{0.3\linewidth}
        \raggedright
        \includegraphics[width=0.9\linewidth]{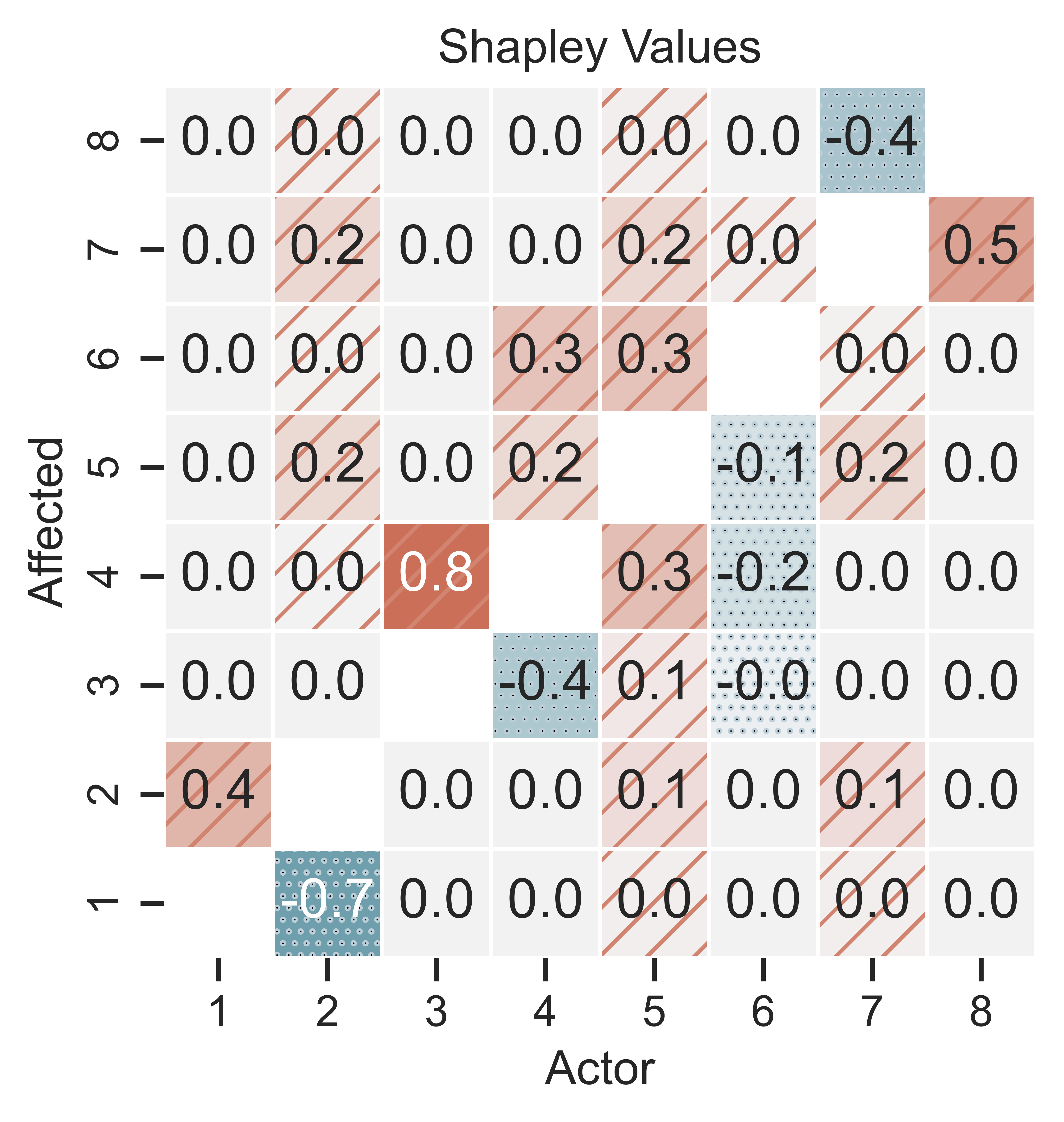}
        \vspace{-0.15cm}
        \caption{\footnotesize Shapley values}
        \label{fig:CrossyRoad-shap}
    \end{subfigure}
    \end{subfigure}
    \begin{subfigure}[c]{0.9\linewidth}
        \begin{subfigure}[c]{0.2\linewidth}
        \vspace{-0.5cm}
        \footnotesize
\begin{tikzpicture}[
    affected/.style={circle, draw=Dandelion!70, fill=white, ultra thick, minimum size=0.6cm, font=\scriptsize},
    agent/.style={circle, draw=red!70, fill=white, very thick, minimum size=0.4cm, font=\scriptsize},
    group/.style={rectangle, rounded corners=10pt, draw=black!40, fill=black!5, very thick, inner sep=3pt},
    tier/.style={rectangle, draw=black!30, fill=black!10, inner sep=5pt},
    arrow/.style={->, >=stealth, ultra thick, draw=black!40, line cap=round},
    link/.style={-, >=stealth, ultra thick, draw=black!40, line cap=round},
]

    % Layer 1: Agent nodes (top)
    \node[affected] (AFFECTED) at (-1.1,0) {1};
    \node[agent] (T0G0A0) at (0,0.4) {5};
    \node[agent] (T0G0A1) at (0,-0.4) {7};

    \begin{pgfonlayer}{background}
        % Tier boundaries (bottom)
        \node[tier, fit=(T0G0A0)(T0G0A1), label={[anchor=south]above:$\tier{1}{1}$ }] (tier0) {};

    % Layer 2: Group boundaries (middle)
        \node[group, fit=(T0G0A0)(T0G0A1)] (group0_0) {};
    \end{pgfonlayer}

    % Layer 4: Arrows (top)
    \draw[arrow] (group0_0) -- (AFFECTED);
\end{tikzpicture}
\normalsize
        \end{subfigure}
        \hfill
        \begin{subfigure}[c]{0.2\linewidth}
        \vspace{-0.5cm}
        \footnotesize
\begin{tikzpicture}[
    affected/.style={circle, draw=Dandelion!70, fill=white, ultra thick, minimum size=0.6cm, font=\scriptsize},
    agent/.style={circle, draw=red!70, fill=white, very thick, minimum size=0.4cm, font=\scriptsize},
    group/.style={rectangle, rounded corners=10pt, draw=black!40, fill=black!5, very thick, inner sep=3pt},
    tier/.style={rectangle, draw=black!30, fill=black!10, inner sep=5pt},
    arrow/.style={->, >=stealth, ultra thick, draw=black!40, line cap=round},
    link/.style={-, >=stealth, ultra thick, draw=black!40, line cap=round},
]

    % Layer 1: Agent nodes (top)
    \node[affected] (AFFECTED) at (-1.1,0) {2};
    \node[agent] (T0G0A0) at (0,0.8) {1};
    \node[agent] (T0G1A0) at (0,-0.0) {5};
    \node[agent] (T0G1A1) at (0,-0.8) {7};

    \begin{pgfonlayer}{background}
        % Tier boundaries (bottom)
        \node[tier, fit=(T0G0A0)(T0G1A0)(T0G1A1), label={[anchor=south]above:$\tier{2}{1}$ }] (tier0) {};

    % Layer 2: Group boundaries (middle)
        \node[group, fit=(T0G1A0)(T0G1A1)] (group0_1) {};
    \end{pgfonlayer}

    % Layer 4: Arrows (top)
    \draw[arrow] (T0G0A0) -- (AFFECTED);
    \draw[arrow] (group0_1) -- (AFFECTED);
\end{tikzpicture}
\normalsize
        \end{subfigure}
        \hfill
        \begin{subfigure}[c]{0.2\linewidth}
        \vspace{-0.5cm}
        \footnotesize
\begin{tikzpicture}[
    affected/.style={circle, draw=Dandelion!70, fill=white, ultra thick, minimum size=0.6cm, font=\scriptsize},
    agent/.style={circle, draw=red!70, fill=white, very thick, minimum size=0.4cm, font=\scriptsize},
    group/.style={rectangle, rounded corners=10pt, draw=black!40, fill=black!5, very thick, inner sep=3pt},
    tier/.style={rectangle, draw=black!30, fill=black!10, inner sep=5pt},
    arrow/.style={->, >=stealth, ultra thick, draw=black!40, line cap=round},
    link/.style={-, >=stealth, ultra thick, draw=black!40, line cap=round},
]

    % Layer 1: Agent nodes (top)
    \node[affected] (AFFECTED) at (-1.1,0) {3};
    \node[agent] (T0G0A0) at (0,0.0) {5};

    \begin{pgfonlayer}{background}
        % Tier boundaries (bottom)
        \node[tier, fit=(T0G0A0), label={[anchor=south]above:$\tier{3}{1}$ }] (tier0) {};

    % Layer 2: Group boundaries (middle)
    \end{pgfonlayer}

    % Layer 4: Arrows (top)
    \draw[arrow] (T0G0A0) -- (AFFECTED);
\end{tikzpicture}
\normalsize
        \end{subfigure}
        \hfill
        \begin{subfigure}[c]{0.3\linewidth}
        \vspace{-0.4cm}
        \footnotesize
\begin{tikzpicture}[
    affected/.style={circle, draw=Dandelion!70, fill=white, ultra thick, minimum size=0.6cm, font=\scriptsize},
    agent/.style={circle, draw=red!70, fill=white, very thick, minimum size=0.4cm, font=\scriptsize},
    group/.style={rectangle, rounded corners=10pt, draw=black!40, fill=black!5, very thick, inner sep=3pt},
    tier/.style={rectangle, draw=black!30, fill=black!10, inner sep=5pt},
    arrow/.style={->, >=stealth, ultra thick, draw=black!40, line cap=round},
    link/.style={-, >=stealth, ultra thick, draw=black!40, line cap=round},
]

    % Layer 1: Agent nodes (top)
    \node[affected] (AFFECTED) at (-1.1,0) {4};
    \node[agent] (T0G0A0) at (0,0.4) {5};
    \node[agent] (T0G1A0) at (0,-0.4) {3};
    \node[agent] (T1G0A0) at (1.1,0.0) {2};

    \begin{pgfonlayer}{background}
        % Tier boundaries (bottom)
        \node[tier, fit=(T0G0A0)(T0G1A0), label={[anchor=south]above:$\tier{4}{1}$ }] (tier0) {};
        \node[tier, fit=(T1G0A0), label={[anchor=south]above:$\tier{4}{2}$ }] (tier1) {};

    % Layer 2: Group boundaries (middle)
    \end{pgfonlayer}

    % Layer 4: Arrows (top)
    \draw[link] (T1G0A0) -- (tier0);
    \draw[arrow] (T0G0A0) -- (AFFECTED);
    \draw[arrow] (T0G1A0) -- (AFFECTED);
\end{tikzpicture}
\normalsize
        \end{subfigure}
        \begin{subfigure}[c]{0.15\linewidth}
        \vspace{-0.9cm}
        \footnotesize
\begin{tikzpicture}[
    affected/.style={circle, draw=Dandelion!70, fill=white, ultra thick, minimum size=0.6cm, font=\scriptsize},
    agent/.style={circle, draw=red!70, fill=white, very thick, minimum size=0.4cm, font=\scriptsize},
    group/.style={rectangle, rounded corners=10pt, draw=black!40, fill=black!5, very thick, inner sep=3pt},
    tier/.style={rectangle, draw=black!30, fill=black!10, inner sep=5pt},
    arrow/.style={->, >=stealth, ultra thick, draw=black!40, line cap=round},
    link/.style={-, >=stealth, ultra thick, draw=black!40, line cap=round},
]

    % Layer 1: Agent nodes (top)
    \node[affected] (AFFECTED) at (-1.1,0) {5};
    \node[agent] (T0G0A0) at (0,0.8) {4};
    \node[agent] (T0G1A0) at (0,-0.0) {2};
    \node[agent] (T0G1A1) at (0,-0.8) {7};

    \begin{pgfonlayer}{background}
        % Tier boundaries (bottom)
        \node[tier, fit=(T0G0A0)(T0G1A0)(T0G1A1), label={[anchor=south]above:$\tier{5}{1}$ }] (tier0) {};

    % Layer 2: Group boundaries (middle)
        \node[group, fit=(T0G1A0)(T0G1A1)] (group0_1) {};
    \end{pgfonlayer}

    % Layer 4: Arrows (top)
    \draw[arrow] (T0G0A0) -- (AFFECTED);
    \draw[arrow] (group0_1) -- (AFFECTED);
\end{tikzpicture}
\normalsize
        \end{subfigure}
        \hfill
        \begin{subfigure}[c]{0.25\linewidth}
        \vspace{-0.9cm}
        \footnotesize
\begin{tikzpicture}[
    affected/.style={circle, draw=Dandelion!70, fill=white, ultra thick, minimum size=0.6cm, font=\scriptsize},
    agent/.style={circle, draw=red!70, fill=white, very thick, minimum size=0.4cm, font=\scriptsize},
    group/.style={rectangle, rounded corners=10pt, draw=black!40, fill=black!5, very thick, inner sep=3pt},
    tier/.style={rectangle, draw=black!30, fill=black!10, inner sep=5pt},
    arrow/.style={->, >=stealth, ultra thick, draw=black!40, line cap=round},
    link/.style={-, >=stealth, ultra thick, draw=black!40, line cap=round},
]

    % Layer 1: Agent nodes (top)
    \node[affected] (AFFECTED) at (-1.1,0) {6};
    \node[agent] (T0G0A0) at (0,0.4) {5};
    \node[agent] (T0G1A0) at (0,-0.4) {4};
    \node[agent] (T1G0A0) at (1.1,0.4) {2};
    \node[agent] (T1G0A1) at (1.1,-0.4) {7};

    \begin{pgfonlayer}{background}
        % Tier boundaries (bottom)
        \node[tier, fit=(T0G0A0)(T0G1A0), label={[anchor=south]above:$\tier{6}{1}$ }] (tier0) {};
        \node[tier, fit=(T1G0A0)(T1G0A1), label={[anchor=south]above:$\tier{6}{2}$ }] (tier1) {};

    % Layer 2: Group boundaries (middle)
        \node[group, fit=(T1G0A0)(T1G0A1)] (group1_0) {};
    \end{pgfonlayer}

    % Layer 4: Arrows (top)
    \draw[link] (group1_0) -- (tier0);
    \draw[arrow] (T0G0A0) -- (AFFECTED);
    \draw[arrow] (T0G1A0) -- (AFFECTED);
\end{tikzpicture}
\normalsize
        \end{subfigure}
        \hfill
        \begin{subfigure}[c]{0.25\linewidth}
        \vspace{-0.9cm}
        \footnotesize
\begin{tikzpicture}[
    affected/.style={circle, draw=Dandelion!70, fill=white, ultra thick, minimum size=0.6cm, font=\scriptsize},
    agent/.style={circle, draw=red!70, fill=white, very thick, minimum size=0.4cm, font=\scriptsize},
    group/.style={rectangle, rounded corners=10pt, draw=black!40, fill=black!5, very thick, inner sep=3pt},
    tier/.style={rectangle, draw=black!30, fill=black!10, inner sep=5pt},
    arrow/.style={->, >=stealth, ultra thick, draw=black!40, line cap=round},
    link/.style={-, >=stealth, ultra thick, draw=black!40, line cap=round},
]

    % Layer 1: Agent nodes (top)
    \node[affected] (AFFECTED) at (-1.1,0) {7};
    \node[agent] (T0G0A0) at (0,0.8) {8};
    \node[agent] (T0G1A0) at (0,-0.0) {2};
    \node[agent] (T0G1A1) at (0,-0.8) {5};
    \node[agent] (T1G0A0) at (1.1,0.0) {6};

    \begin{pgfonlayer}{background}
        % Tier boundaries (bottom)
        \node[tier, fit=(T0G0A0)(T0G1A0)(T0G1A1), label={[anchor=south]above:$\tier{7}{1}$ }] (tier0) {};
        \node[tier, fit=(T1G0A0), label={[anchor=south]above:$\tier{7}{2}$ }] (tier1) {};

    % Layer 2: Group boundaries (middle)
        \node[group, fit=(T0G1A0)(T0G1A1)] (group0_1) {};
    \end{pgfonlayer}

    % Layer 4: Arrows (top)
    \draw[link] (T1G0A0) -- (tier0);
    \draw[arrow] (T0G0A0) -- (AFFECTED);
    \draw[arrow] (group0_1) -- (AFFECTED);
\end{tikzpicture}
\normalsize
        \end{subfigure}
        \hfill
        \begin{subfigure}[c]{0.2\linewidth}
        \vspace{-0.9cm}
        \footnotesize
\begin{tikzpicture}[
    affected/.style={circle, draw=Dandelion!70, fill=white, ultra thick, minimum size=0.6cm, font=\scriptsize},
    agent/.style={circle, draw=red!70, fill=white, very thick, minimum size=0.4cm, font=\scriptsize},
    group/.style={rectangle, rounded corners=10pt, draw=black!40, fill=black!5, very thick, inner sep=3pt},
    tier/.style={rectangle, draw=black!30, fill=black!10, inner sep=5pt},
    arrow/.style={->, >=stealth, ultra thick, draw=black!40, line cap=round},
    link/.style={-, >=stealth, ultra thick, draw=black!40, line cap=round},
]

    % Layer 1: Agent nodes (top)
    \node[affected] (AFFECTED) at (-1.1,0) {8};
    \node[agent] (T0G0A0) at (0,0.4) {2};
    \node[agent] (T0G0A1) at (0,-0.4) {5};

    \begin{pgfonlayer}{background}
        % Tier boundaries (bottom)
        \node[tier, fit=(T0G0A0)(T0G0A1), label={[anchor=south]above:$\tier{8}{1}$ }] (tier0) {};

    % Layer 2: Group boundaries (middle)
        \node[group, fit=(T0G0A0)(T0G0A1)] (group0_0) {};
    \end{pgfonlayer}

    % Layer 4: Arrows (top)
    \draw[arrow] (group0_0) -- (AFFECTED);
\end{tikzpicture}
\normalsize
        \end{subfigure}
        \vspace{-0.8cm}
        \caption{\footnotesize Tiers affecting different agents from group FeAR.}
        \label{fig:CrossyRoad-Tiers}
    \end{subfigure}
    \caption{
    \small
    \textbf{S1: Uncovering group effects with group FeAR:}
    For the robot crossing scenario from \cref{fig:CrossyRoadIntro}, represented in the grid world as in (a), 
    compared to just using iFeAR~(b),
    we are able to uncover more assertive influences using gFeAR, either through
    the Shapley values~(c) or tiers~(d). 
    For example, while iFeAR only shows the assertive influence of the robot 5 on pedestrians 3, 4 and 6, 
    both Shapley values and tiers show that 5 is assertive towards all the pedestrians.
    }
    \label{fig:CrossyRoadResults}
\end{figure}
\normalsize

\begin{table}[htb]
\centering
    \caption{
    \small
    \textbf{Ranks of assertive influence in S1:} Ranking of assertive influence calculated using FeAR, tiers from group FeAR and Shapley values of group FeAR. `\textcolor{lightgray}{$\blacksquare$}' show affected agents which are not considered in the rankings.}
    \label{tab:CrossyRoadResults}
    \scriptsize
    \begin{subtable}[c]{\linewidth}
    \caption{Ranking's of assertiveness and $\deltaAssertive$.}
        \label{tab:rankings_crossyRoad}
    \begin{subtable}[t]{0.5\linewidth}
        \centering
             \begin{tabular}{r l r}
             %\toprule
             %\ & \textbf{Actors} & : $\begin{bmatrix} 1 & 2 & 3 & 4 & 5 & 6 & 7 & 8 ~\end{bmatrix}$\\
             %\midrule
             \multicolumn{2}{l}{\textbf{Affected: 1 } } & $\deltaAssertive=2$ \,\\
             \midrule
             iFeAR &           ranks & : $\begin{bmatrix} \color{lightgray}{\blacksquare} & \color{lightgray}{\phdot} & \color{lightgray}{\phdot} & \color{lightgray}{\phdot} & \color{lightgray}{\phdot} & \color{lightgray}{\phdot} & \color{lightgray}{\phdot} & \color{lightgray}{\phdot} \end{bmatrix}$\\
             gFeAR & (Tier)    ranks & : $\begin{bmatrix} \color{lightgray}{\blacksquare} & \color{lightgray}{\phdot} & \color{lightgray}{\phdot} & \color{lightgray}{\phdot} & 1 & \color{lightgray}{\phdot} & 1 & \color{lightgray}{\phdot} \end{bmatrix}$\\
             gFeAR & (Shapley) ranks & : $\begin{bmatrix} \color{lightgray}{\blacksquare} & \color{lightgray}{\phdot} & \color{lightgray}{\phdot} & \color{lightgray}{\phdot} & 1 & \color{lightgray}{\phdot} & 1 & \color{lightgray}{\phdot} \end{bmatrix}$\\
\bottomrule
\end{tabular}
    \end{subtable}
    \vspace{0.1cm}
    \begin{subtable}[t]{0.5\linewidth}
        \centering
             \begin{tabular}{r l r}
             %\toprule
             %\ & \textbf{Actors} & : $\begin{bmatrix} 1 & 2 & 3 & 4 & 5 & 6 & 7 & 8 ~\end{bmatrix}$\\
             %\midrule
             \multicolumn{2}{l}{\textbf{Affected: 2 } } & $\deltaAssertive=2$ \,\\
             \midrule
             iFeAR &           ranks & : $\begin{bmatrix} 1 & \color{lightgray}{\blacksquare} & \color{lightgray}{\phdot} & \color{lightgray}{\phdot} & \color{lightgray}{\phdot} & \color{lightgray}{\phdot} & \color{lightgray}{\phdot} & \color{lightgray}{\phdot} \end{bmatrix}$\\
             gFeAR & (Tier)    ranks & : $\begin{bmatrix} 1 & \color{lightgray}{\blacksquare} & \color{lightgray}{\phdot} & \color{lightgray}{\phdot} & 1 & \color{lightgray}{\phdot} & 1 & \color{lightgray}{\phdot} \end{bmatrix}$\\
             gFeAR & (Shapley) ranks & : $\begin{bmatrix} 1 & \color{lightgray}{\blacksquare} & \color{lightgray}{\phdot} & \color{lightgray}{\phdot} & 2 & \color{lightgray}{\phdot} & 2 & \color{lightgray}{\phdot} \end{bmatrix}$\\
\bottomrule
\end{tabular}
    \end{subtable}
    \vspace{0.1cm}
    \begin{subtable}[t]{0.5\linewidth}
        \centering
             \begin{tabular}{r l r}
             %\toprule
             %\ & \textbf{Actors} & : $\begin{bmatrix} 1 & 2 & 3 & 4 & 5 & 6 & 7 & 8 ~\end{bmatrix}$\\
             %\midrule
             \multicolumn{2}{l}{\textbf{Affected: 3 } } & $\deltaAssertive=0$ \,\\
             \midrule
             iFeAR &           ranks & : $\begin{bmatrix} \color{lightgray}{\phdot} & \color{lightgray}{\phdot} & \color{lightgray}{\blacksquare} & \color{lightgray}{\phdot} & 1 & \color{lightgray}{\phdot} & \color{lightgray}{\phdot} & \color{lightgray}{\phdot} \end{bmatrix}$\\
             gFeAR & (Tier)    ranks & : $\begin{bmatrix} \color{lightgray}{\phdot} & \color{lightgray}{\phdot} & \color{lightgray}{\blacksquare} & \color{lightgray}{\phdot} & 1 & \color{lightgray}{\phdot} & \color{lightgray}{\phdot} & \color{lightgray}{\phdot} \end{bmatrix}$\\
             gFeAR & (Shapley) ranks & : $\begin{bmatrix} \color{lightgray}{\phdot} & \color{lightgray}{\phdot} & \color{lightgray}{\blacksquare} & \color{lightgray}{\phdot} & 1 & \color{lightgray}{\phdot} & \color{lightgray}{\phdot} & \color{lightgray}{\phdot} \end{bmatrix}$\\
\bottomrule
\end{tabular}
    \end{subtable}
    \vspace{0.1cm}
    \begin{subtable}[t]{0.5\linewidth}
        \centering
             \begin{tabular}{r l r}
             %\toprule
             %\ & \textbf{Actors} & : $\begin{bmatrix} 1 & 2 & 3 & 4 & 5 & 6 & 7 & 8 ~\end{bmatrix}$\\
             %\midrule
             \multicolumn{2}{l}{\textbf{Affected: 4 } } & $\deltaAssertive=1$ \,\\
             \midrule
             iFeAR &           ranks & : $\begin{bmatrix} \color{lightgray}{\phdot} & \color{lightgray}{\phdot} & 1 & \color{lightgray}{\blacksquare} & 1 & \color{lightgray}{\phdot} & \color{lightgray}{\phdot} & \color{lightgray}{\phdot} \end{bmatrix}$\\
             gFeAR & (Tier)    ranks & : $\begin{bmatrix} \color{lightgray}{\phdot} & 3 & 1 & \color{lightgray}{\blacksquare} & 1 & \color{lightgray}{\phdot} & \color{lightgray}{\phdot} & \color{lightgray}{\phdot} \end{bmatrix}$\\
             gFeAR & (Shapley) ranks & : $\begin{bmatrix} \color{lightgray}{\phdot} & 3 & 1 & \color{lightgray}{\blacksquare} & 2 & \color{lightgray}{\phdot} & \color{lightgray}{\phdot} & \color{lightgray}{\phdot} \end{bmatrix}$\\
\bottomrule
\end{tabular}
    \end{subtable}
    \vspace{0.1cm}
    \begin{subtable}[t]{0.5\linewidth}
        \centering
             \begin{tabular}{r l r}
             %\toprule
             %\ & \textbf{Actors} & : $\begin{bmatrix} 1 & 2 & 3 & 4 & 5 & 6 & 7 & 8 ~\end{bmatrix}$\\
             %\midrule
             \multicolumn{2}{l}{\textbf{Affected: 5 } } & $\deltaAssertive=2$ \,\\
             \midrule
             iFeAR &           ranks & : $\begin{bmatrix} \color{lightgray}{\phdot} & \color{lightgray}{\phdot} & \color{lightgray}{\phdot} & 1 & \color{lightgray}{\blacksquare} & \color{lightgray}{\phdot} & \color{lightgray}{\phdot} & \color{lightgray}{\phdot} \end{bmatrix}$\\
             gFeAR & (Tier)    ranks & : $\begin{bmatrix} \color{lightgray}{\phdot} & 1 & \color{lightgray}{\phdot} & 1 & \color{lightgray}{\blacksquare} & \color{lightgray}{\phdot} & 1 & \color{lightgray}{\phdot} \end{bmatrix}$\\
             gFeAR & (Shapley) ranks & : $\begin{bmatrix} \color{lightgray}{\phdot} & 1 & \color{lightgray}{\phdot} & 3 & \color{lightgray}{\blacksquare} & \color{lightgray}{\phdot} & 1 & \color{lightgray}{\phdot} \end{bmatrix}$\\
\bottomrule
\end{tabular}
    \end{subtable}
    \vspace{0.1cm}
    \begin{subtable}[t]{0.5\linewidth}
        \centering
             \begin{tabular}{r l r}
             %\toprule
             %\ & \textbf{Actors} & : $\begin{bmatrix} 1 & 2 & 3 & 4 & 5 & 6 & 7 & 8 ~\end{bmatrix}$\\
             %\midrule
             \multicolumn{2}{l}{\textbf{Affected: 6 } } & $\deltaAssertive=2$ \,\\
             \midrule
             iFeAR &           ranks & : $\begin{bmatrix} \color{lightgray}{\phdot} & \color{lightgray}{\phdot} & \color{lightgray}{\phdot} & 1 & 1 & \color{lightgray}{\blacksquare} & \color{lightgray}{\phdot} & \color{lightgray}{\phdot} \end{bmatrix}$\\
             gFeAR & (Tier)    ranks & : $\begin{bmatrix} \color{lightgray}{\phdot} & 3 & \color{lightgray}{\phdot} & 1 & 1 & \color{lightgray}{\blacksquare} & 3 & \color{lightgray}{\phdot} \end{bmatrix}$\\
             gFeAR & (Shapley) ranks & : $\begin{bmatrix} \color{lightgray}{\phdot} & 3 & \color{lightgray}{\phdot} & 1 & 1 & \color{lightgray}{\blacksquare} & 3 & \color{lightgray}{\phdot} \end{bmatrix}$\\
\bottomrule
\end{tabular}
    \end{subtable}
    \vspace{0.1cm}
    \begin{subtable}[t]{0.5\linewidth}
        \centering
             \begin{tabular}{r l r}
             %\toprule
             %\ & \textbf{Actors} & : $\begin{bmatrix} 1 & 2 & 3 & 4 & 5 & 6 & 7 & 8 ~\end{bmatrix}$\\
             %\midrule
             \multicolumn{2}{l}{\textbf{Affected: 7 } } & $\deltaAssertive=3$ \,\\
             \midrule
             iFeAR &           ranks & : $\begin{bmatrix} \color{lightgray}{\phdot} & \color{lightgray}{\phdot} & \color{lightgray}{\phdot} & \color{lightgray}{\phdot} & \color{lightgray}{\phdot} & \color{lightgray}{\phdot} & \color{lightgray}{\blacksquare} & 1 \end{bmatrix}$\\
             gFeAR & (Tier)    ranks & : $\begin{bmatrix} \color{lightgray}{\phdot} & 1 & \color{lightgray}{\phdot} & \color{lightgray}{\phdot} & 1 & 4 & \color{lightgray}{\blacksquare} & 1 \end{bmatrix}$\\
             gFeAR & (Shapley) ranks & : $\begin{bmatrix} \color{lightgray}{\phdot} & 2 & \color{lightgray}{\phdot} & \color{lightgray}{\phdot} & 2 & 4 & \color{lightgray}{\blacksquare} & 1 \end{bmatrix}$\\
\bottomrule
\end{tabular}
    \end{subtable}
    \vspace{0.1cm}
    \begin{subtable}[t]{0.5\linewidth}
        \centering
             \begin{tabular}{r l r}
             %\toprule
             %\ & \textbf{Actors} & : $\begin{bmatrix} 1 & 2 & 3 & 4 & 5 & 6 & 7 & 8 ~\end{bmatrix}$\\
             %\midrule
             \multicolumn{2}{l}{\textbf{Affected: 8 } } & $\deltaAssertive=2$ \,\\
             \midrule
             iFeAR &           ranks & : $\begin{bmatrix} \color{lightgray}{\phdot} & \color{lightgray}{\phdot} & \color{lightgray}{\phdot} & \color{lightgray}{\phdot} & \color{lightgray}{\phdot} & \color{lightgray}{\phdot} & \color{lightgray}{\phdot} & \color{lightgray}{\blacksquare} \end{bmatrix}$\\
             gFeAR & (Tier)    ranks & : $\begin{bmatrix} \color{lightgray}{\phdot} & 1 & \color{lightgray}{\phdot} & \color{lightgray}{\phdot} & 1 & \color{lightgray}{\phdot} & \color{lightgray}{\phdot} & \color{lightgray}{\blacksquare} \end{bmatrix}$\\
             gFeAR & (Shapley) ranks & : $\begin{bmatrix} \color{lightgray}{\phdot} & 1 & \color{lightgray}{\phdot} & \color{lightgray}{\phdot} & 1 & \color{lightgray}{\phdot} & \color{lightgray}{\phdot} & \color{lightgray}{\blacksquare} \end{bmatrix}$\\
\bottomrule
\end{tabular}
    \end{subtable}
    \end{subtable}
    \begin{subtable}{\linewidth}
    \centering
        \caption{Kendall's $\tau$ comparing rankings. $\tau$ close to 1 indicate similar rankings.}
        \label{tab:taus_crossyRoad}
        {
        \setlength{\tabcolsep}{4pt}
        \begin{tabular}{lcccccccc}
        \toprule
Affected & 1 & 2 & 3 & 4 & 5 & 6 & 7 & 8 \\
        \midrule
$\tau$(iFeAR, gFeAR-(Tier)) & - & 0.47 & 1.00 & 0.85 & 0.47 & 0.79 & 0.42 & - \\
$\tau$(iFeAR, gFeAR-(Shapley)) & - & 0.65 & 1.00 & 0.82 & 0.22 & 0.79 & 0.59 & - \\
$\tau$(gFeAR-(Tier), gFeAR-(Shapley)) & 1.00 & 0.93 & 1.00 & 0.97 & 0.93 & 1.00 & 0.94 & 1.00 \\
        \bottomrule
        \end{tabular}

        }
    \end{subtable}

\normalsize
\end{table}

\section{Results}
\label{sec:Results}

Groups effects in four scenarios were analysed with respect to the difference in the number of assertive agents identified and Kendall's $\tau$ for comparing rankings of assertiveness.

\subsection{Number of assertive agents from individual and group FeAR}
\label{sec:Results:n-assertive}

The tiers for scenario S1 shown in \cref{tab:CrossyRoadResults} show how the robot (agent 5) is affected by the solo influence of pedestrian 4 ($\solo{4}{5}$) and coupled influence of pedestrians 2 and 7 ($\coupled{\{2,7\}}{5}$), and how it affects all the pedestrians either through solo (
$\solo{5}{3}$,
$\solo{5}{4}$,
$\solo{5}{6}$
) or coupled influence (
$\coupled{\{5,7\}}{1}$,
$\coupled{\{5,7\}}{2}$,
$\coupled{\{2,5\}}{7}$,
$\coupled{\{2,5\}}{8}$ 
). 
Besides these influences involving agent 5, the algorithm also identifies mediated influences (
$\mediated{2}{4}{\tier{4}{1}}$,
$\mediated{6}{7}{\tier{7}{1}}$
) and mediated coupled influence ($\mediatedCoupled{\{2,7\}}{6}{\tier{6}{1}}$) which are mediated by tiers $\tier{j}{1}$ of affected agents $j$.

Thus, by identifying coupled and mediated influences, group FeAR identifies more agents which are being assertive towards an affected agent, than would be possible just by using individual FeAR (as shown by the values of $\deltaAssertive=\nAssertivegFeAR-\nAssertiveiFeAR$ in \cref{tab:CrossyRoadResults}). 
Since iFeAR can only identify solo influences, it can never identify more assertive agents than the tiers from gFeAR; so
$\nAssertivegFeAR \geq \nAssertiveiFeAR$ and $\deltaAssertive \geq 0$. 

In S1, the value of $\deltaAssertive$ has a maximum of three for affected agent 7. For agent 7, in addition to the solo influence of agent 8 ($
\solo{7}{8}$), the tiers reveal the coupled influence of agents 2 and 5 ($\coupled{7}{\group{2,5}}$), and the mediated influence of agent 6 ($\mediated{7}{6}{\group{2,5,8}}$). 

Furthermore, agents 6 and 7 have the maximum number of assertive agents ($\nAssertivegFeAR=4$) followed by agents 2, 4 and 5 ($\nAssertivegFeAR=3$). These agents are in the centre of the interaction and their proximity to other agents might be a contributing factor for the group effects on them. 

\begin{figure}
    \centering
\begin{subfigure}[c]{\linewidth}
\centering
\begin{subfigure}[t]{0.2\linewidth}
    \centering
    \inColouredBox{Magenta}{S2:Aggressive}{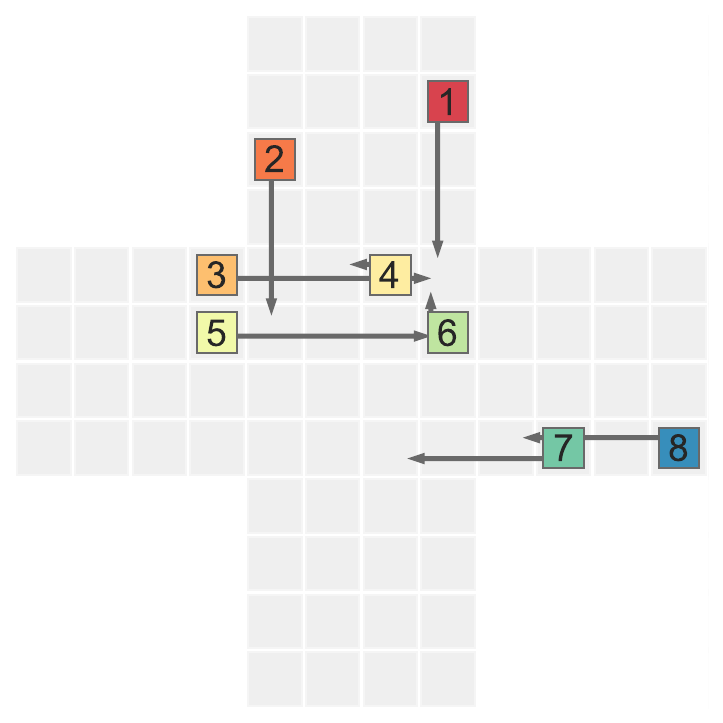}
\end{subfigure}
\hspace{1cm}
\begin{subfigure}[t]{0.2\linewidth}
    \centering
    \inColouredBox{Cyan}{S3:Directed}{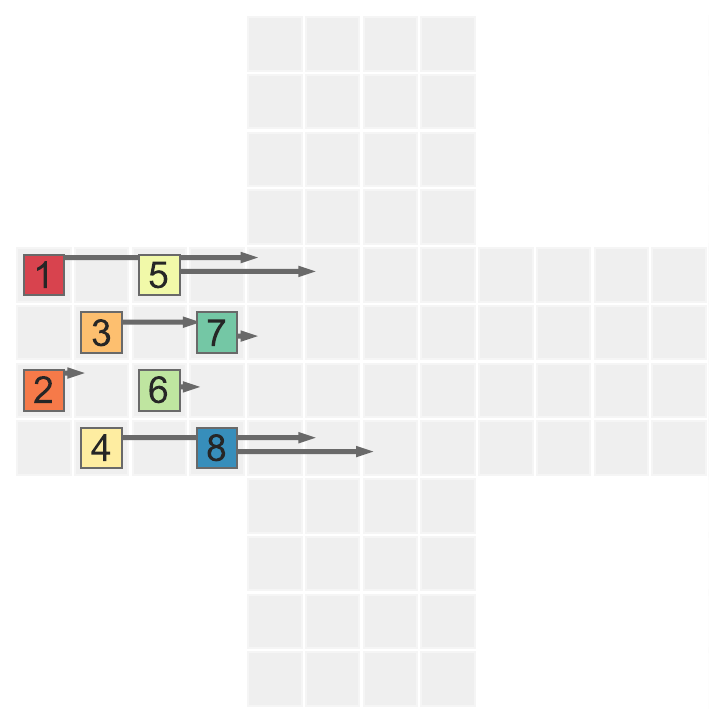}
\end{subfigure}
\hspace{1cm}
\begin{subfigure}[t]{0.2\linewidth}
    \centering
    \inColouredBox{Goldenrod}{S4:Random}{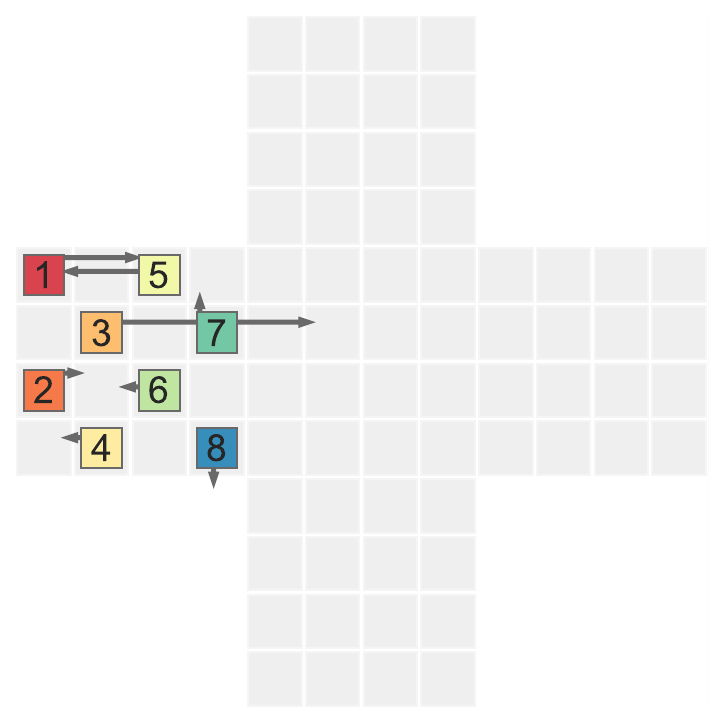}
\end{subfigure}
\vspace{-0.3cm}
\caption{Scenarios}
\label{fig:sims-scenarios}
\end{subfigure}

\begin{subfigure}[c]{\linewidth}
\begin{subfigure}[t]{0.57\linewidth}
    \centering
    \includegraphics[width=\linewidth]{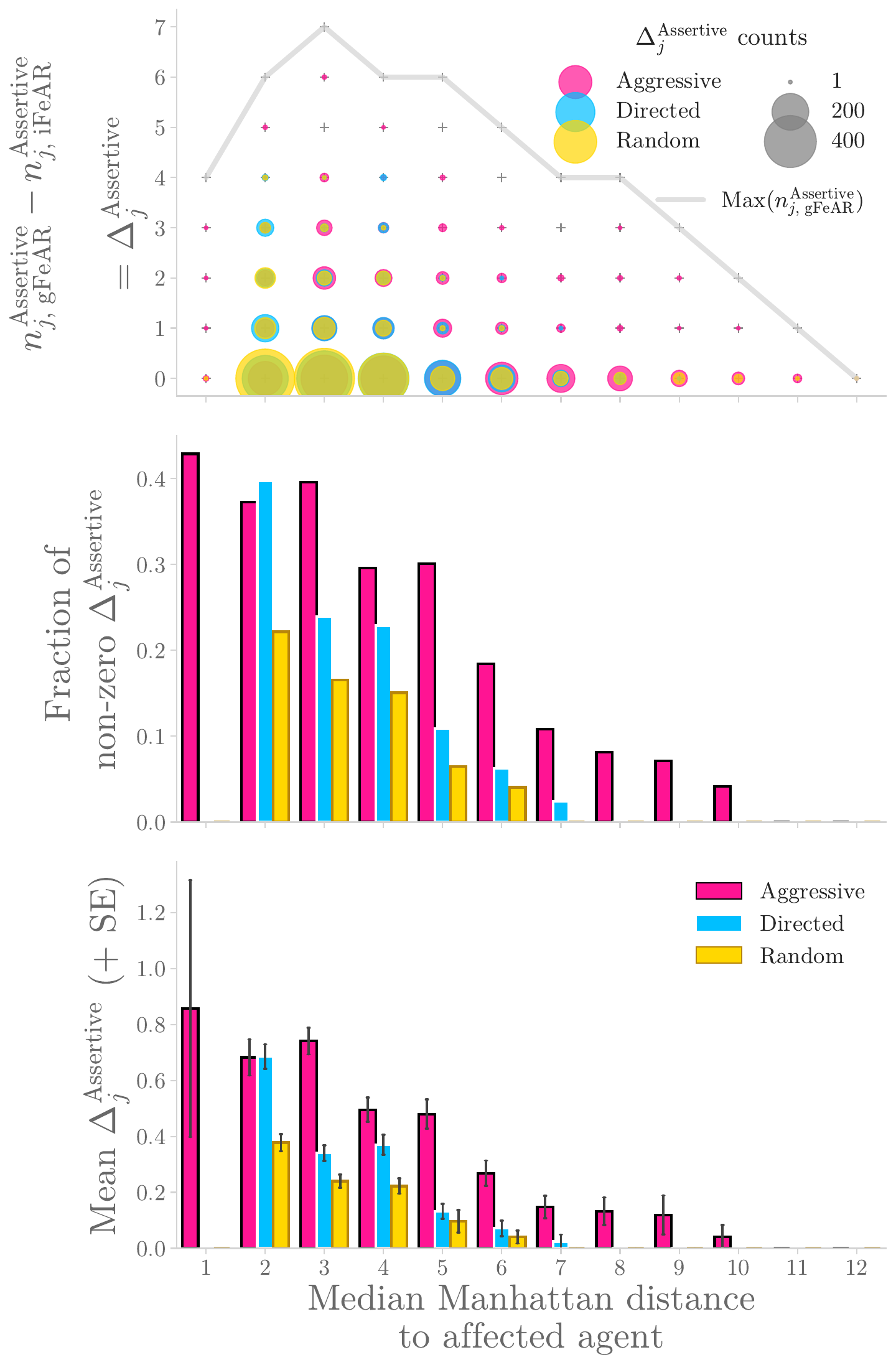}
    \normalsize
    \caption{Difference in the number of assertive agents identified using gFeAR and iFeAR $\deltaAssertive$.}
    \label{fig:n-assertive-tier-fear}
\end{subfigure}
\hfill
\begin{subfigure}[t]{0.4\linewidth}
    \centering
    \includegraphics[width=1\linewidth]{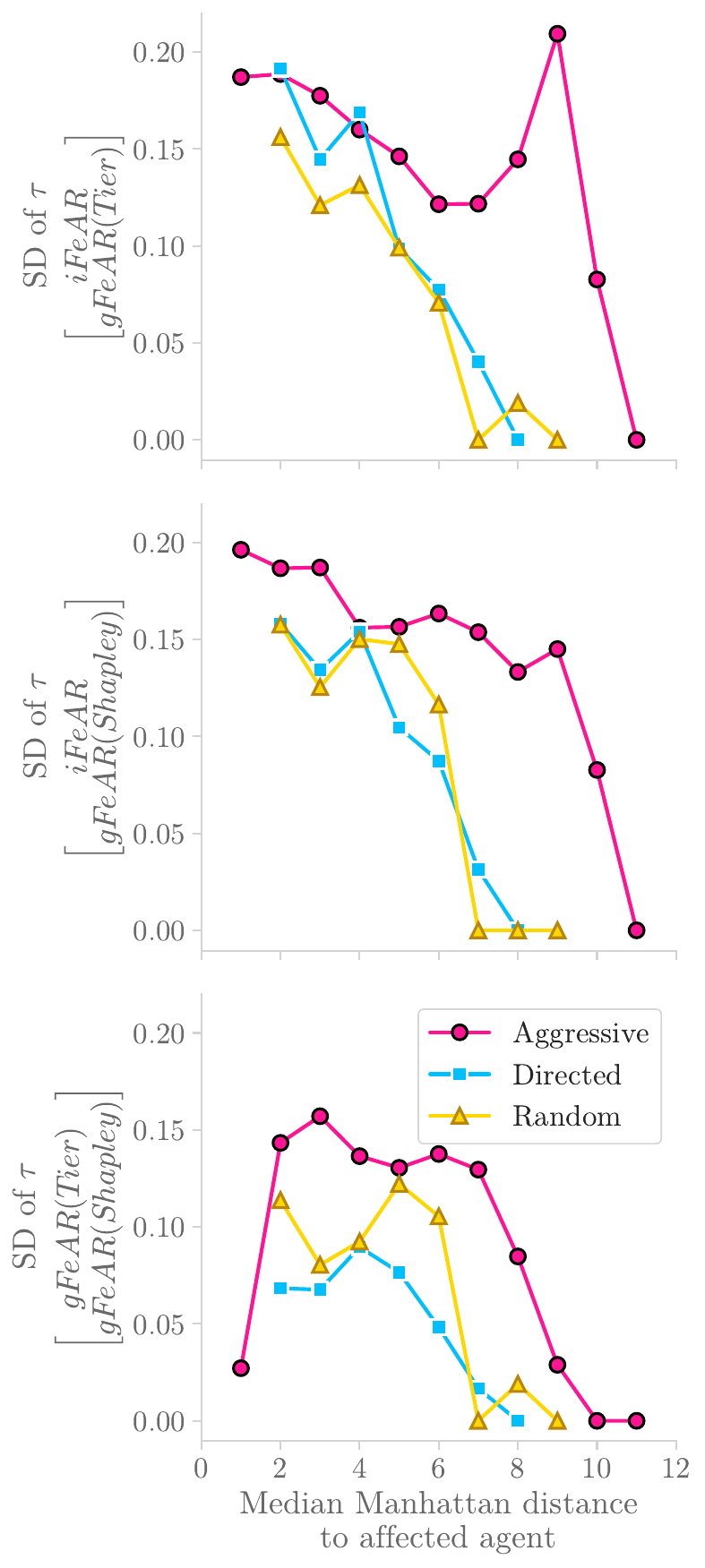}
    \caption{Standard deviation of Kendall's~$\tau$ for comparing ranks.}
    \label{fig:std-tau-vs-manhattan-distance}
\end{subfigure}
\end{subfigure}
\caption{
\small
\textbf{Emergence of group effects in randomised simulations:}
Different simulation scenarios are shown in (a).
Difference in the number of assertive agents $\deltaAssertive$ identified using individual FeAR (iFeAR) and (tiers of) group FeAR (gFeAR) for different proximity to the affected agent are shown in (b).
Rankings made using FeAR, tiers from gFeAR or Shapley values of gFeAR we compared using Kendall's $\tau$.
Variation of Kendall's $\tau$ with respect to the scenarios and median Manhattan distance between agents is shown in (c).
}
\label{fig:RandomisedSimualtions}
\end{figure}

To further explore the effect of proximity on group effects in the randomised simulations, we plot $\deltaAssertive$ versus median Manhattan distance of other agents to the affected agent~(\cref{fig:n-assertive-tier-fear}). Low values of the median Manhattan distance would imply that other agents were more proximal to the affected agent. It should be noted that the median Manhattan distance to affected agent is a meaningful metric only because all the scenarios considered here have the same map and number of agents.
As the median Manhattan distance increases beyond a threshold, the maximum $\nAssertivegFeAR$ line show how the number of assertive agents start to drop. 
Counts of instances of $\deltaAssertive$ for different scenarios also show mirror this trend of decreasing group effects as the affected agents gets farther from other agents. To better understand the trends in group effects we also plot the fraction of non-zero $\deltaAssertive$ and mean $\deltaAssertive$ for each scenario in \cref{fig:n-assertive-tier-fear}. Both of these show how group effects decrease as the proximity of the affected agent to others decrease. 

Another key finding is how group effects vary across different simulation scenarios. The fraction of non-zero $\deltaAssertive$ and mean $\deltaAssertive$ for each mean Manhattan distance show how group effects are generally the largest in case of the \emph{Aggressive} scenario where all the agents are aggressively crossing the intersection.
Also, for larger distances from the affected agent, group effects are still present for \emph{Assertive}, while they die off for \emph{Directed} and \emph{Random}.
Of the three scenarios, \emph{Random} has the lowest values of $\deltaAssertive$ while \emph{Directed} has intermediate values. In summary, the values of $\deltaAssertive$ show that group effects are strongest in \emph{Assertive} and weakest in \emph{Random}.  

\subsection{Comparing rankings of assertiveness}
\label{sec:Results-CompareRanks}

So far, we have explored the difference in the number of assertive agents identified using individual FeAR and group FeAR. Besides just identifying assertive agents, our algorithm ranks assertive influences into tiers. Here, we will compare these $\tierRanks$ ranks against $\fearRanks$ ranks generated from individual FeAR values and $\shapRanks$ ranks generated from Shapley values based in group FeAR.

The rankings for the robot crossing scenario (CS1) are summarised in \cref{tab:CrossyRoadResults} in terms of ranks. So for affected agent 7, the $\tierRanks$ ranks of $\left(1,1,1,4\right)$ for agents 2, 5, 8 and 6 represent the ranking $2\sim5\sim8\succ6$ and the $\shapRanks$ ranks of $\left(1,2,2,4\right)$ for agents 8, 2, 5 and 6, represent the ranking $8\succ2\sim5\succ6$.

Apart from fewer assertive agents identified using iFeAR, most of the other rankings are in agreement with each other. One difference between $\tierRanks$ and $\shapRanks$ ranks concerns the ranking of coupled influences. Coupled influences are consistently ranked along solo influences in $\fearRanks$ ranks, while the $\shapRanks$ ranks for coupled influences is higher than the solo influence for affected agent 5, and lower than solo influences for agents 2 and 7. Another difference between $\shapRanks$ and $\tierRanks$ ranks is in the ranks of solo influence --- agents with solo influence always have rank 1 in $\tierRanks$, whereas $\shapRanks$ ranks for agents with solo influence can be different. For example, in the $\shapRanks$ ranks for affected agent 4, the ranks of agents 3 and 5 are 1 and 2 respectively, whereas both of them have $\tierRanks$ of 1.   

The difference among these rankings were quantified using Kendall's $\tau$ which had values in the range $\left[-1, 1\right]$ and complete agreement among rankings had $\tau=1$. Based on the difference of $\taufearTier$ from 1, the difference between $\fearRanks$ and $\tierRanks$ ranks was largest for agent 7 and smallest for agent 3 ($7>2\sim5>6>4>3$).
It must be noted that even though agents 1 and 8 experience group effects due to coupled influences, the lack of $\fearRanks$ makes computing $\taufearTier$ impossible.
Even with this caveat, the values of $\taufearTier$ give a good indication for the overall presence of group effects in the scenario.
Thus,
the values of $\taufearTier$ show how group effects are stronger on central agents that are more proximal to other agents.
Similarly, the values of $\taufearShap$ which represent the difference between the $\fearRanks$ and $\shapRanks$, 
also highlight the prevalence of groups effects on  central agents --- with $1-\tau$ values for different affected agents ranked as
$5>7>2>6>4>3$.

In addition to comparing iFeAR and gFeAR using $\taufearTier$, we can use $\tauTierShap$ to compare the two rankings generated from gFeAR. All the values of $\tauTierShap<0.9$ indicate strong agreement between $\tierRanks$
and $\shapRanks$ ranks. Again the values of $1-\tau$ for different affected agents ($2\sim5>7>4>1\sim3\sim6\sim8$) indicate greater differences in central agents. 

Thus, since the group effects on an affected agent might be related to its centrality and proximity to other agents, quantified proximity to other agents using the median Manhattan distance and explored how taus were related to this distance in different scenarios. We are interested in the spread of $\tau$ values and hence plot the standard deviation (SD) of $\tau$ in different scenarios for different median Manhattan distance~(\cref{fig:std-tau-vs-manhattan-distance}).

For both $\taufearTier$ and $\taufearShap$ comparing iFeAR and gFeAR, we can see that the SD in $\tau$ drops as the median Manhattan distance to the affected agent increases. Furthermore, we can see that the SD in $\tau$ is mcuh greater for \emph{Aggressive} than for \emph{Directed} or \emph{Random}.

Regarding the values of $\tauTierShap$, the spread of the SDs indicate differences between $\tierRanks$ and $\shapRanks$, but these differences are smaller than their difference with $\fearRanks$ ranks.

\section{Discussion}
\label{sec:Discussion}

% Here, we will discuss the importance of these rankings of assertiveness for filling gaps in causal responsibility~(\cref{sec:CausalResponsibilityGaps}) and for detecting emergence in multi-agent spatial interactions~(\cref{sec:Metric4Emergence}).

\subsection{Filling causal responsibility gaps with Group FeAR}
\label{sec:CausalResponsibilityGaps}

One of the main challenges with AI agents and collective actions is the possibility of responsibility voids ~\cite{matthiasResponsibilityGapAscribing2004,brahamVoidsFragmentationMoral2018,santonidesioFourResponsibilityGaps2021,goetzeMindGapAutonomous2022,veluwenkampWhatResponsibilityGaps2025}. Responsibility voids occur when no one can be held responsible for an outcome that resulted from collective action~\cite{brahamVoidsFragmentationMoral2018,duijfResponsibilityVoidsCooperation2018}. When a group of agents can be held responsible, but no individual can be held responsible, this is called a responsibility gap~\cite{duijfLogicalStudyMoral2023}. The discourse on responsibility voids and responsibility gaps revolve around moral responsibility, which depends on conditions like intention, knowledge, and wrong-doing on top of causally contributing to the outcome. 
In this paper, we skip the all other conditions and focus on causal responsibility and thereby on causal responsibility gaps.

We define that causal responsibility gaps occur when a group of agents have collective causal responsibility, but none of the individuals on their own can be ascribed causal responsibility. Such causal responsibility gaps can occur if we solely rely on the FeAR values for individual agents \cite{georgeFeasibleActionSpaceReduction2023a,georgeFeasibleActionSpace2025}. According to the definition of iFeAR as in \cref{Eq:FeAR}, the causal responsibility of an agent for it's own trajectory is determined as the complement of the FeAR imposed on it by all other agents $(\FeAR_{j,j} = 1-\FeAR_{\neg j,j})$. Thus, cases of coupled influence can lead to $\FeAR_{i,j}=0\, \forall i \in \Agents$, where no individual agent is causally responsible, but $\neg j$ as a collective has assertive influence on $j$. If we were to blindly hold all agents in $\neg j$ responsible, this would lead to 
 responsibility gluts ``where too many agents are held responsible'' \cite{duijfLogicalStudyMoral2023}. To prevent both causal responsibility gaps and gluts, our tiering algorithm for sorting assertive influences, systematically probes the assertiveness of individual agents and groups to identify minimal groups with assertive influence.

Minimality of the sufficiency set is an important criteria when inferring causality, i.e., all elements much be necessary to cause the outcome~\cite{triantafyllouActualCausalityResponsibility2022,halpernModificationHalpernpearlDefinition2015,brahamDegreesCausation2009}. Instead of focussing on a particular outcome, we are interested in a group's degree of causal influence on the trajectory of an affected agent, which can increase with the number of agents within the group. 
Using systematic and incremental interventions grounded on the actual joint action of agents, the tiering algorithm is able to unravel the structure of causal influence on affected agents.

Shapley values have been widely used to compute the contributions of individual agents to collective rewards or costs~\cite{shapleyValueNPersonGames1953}. To compute Shapely values of group FeAR for $k$ agents on an affected agent, we need to consider $2^k$ counterfactual scenarios. By eliminating courteous agents (with $\FeAR_{i,j}<0$) and grouping agents in higher tiers, the tiering algorithm potentially saves on computation cost. After identifying all assertive agents using the tiering algorithm, Shapley values could be useful in distributing penalties to individuals.

In this paper, we compute causal responsibility based on feasible action-space reduction (FeAR) which simply looks at how actions of agents reduce the feasible action-space of others by causing collisions. There are other models of responsibility that are base on game-theoretic formulations~\cite{duijfLogicalStudyMoral2023,gladyshevGroupResponsibilityExceeding2023}, logical formulations of ability and obligations~\cite{loriniLogicalAnalysisResponsibility2014,yazdanpanahDistantGroupResponsibility2016,yazdanpanahApplyingStrategicReasoning2021}, and probabilistic models of causation~\cite{chocklerResponsibilityBlameStructuralModel2004,halpernCauseResponsibilityBlame2015,englTheoryCausalResponsibility2018,alechinaCausalityResponsibilityBlame2020} --- some even consider epistemic states like knowledge and intentions to properly ascribe moral responsibility. Compared to these models of responsibility, FeAR provides a simplistic and model-agnostic metric for causal responsibility, that can be applied to a particular time window of spatial interaction. These simplifications make it an ideal candidate for parsing causal responsibility in large datasets of spatial interactions. Short-listed scenarios with (problematic) causal responsibility ascriptions can be further subjected to more rigorous scrutiny with regard to epistemic, probabilistic and motivational characteristics.

\subsection{Metric for emergence in spatial interactions}
\label{sec:Metric4Emergence}
In our simulations there was little room for emergence as agents did not learn or interact with each other. But by prescribing top-down policies we were hoping to generate ``emergence-like'' behaviour. In doing so, we have stumbled on some metrics that might be useful in flagging potential emergent behaviours in spatial interactions.

Emergence can be defined as ``the appearance of patterns, properties and behaviours within a system that are not evident in individual components~\cite{greenEmergenceComplexNetworks2023}.''
Based on the relationship between the macroscopic property and microscopic factors that cause it, emergence has been classified into \emph{nominal}, \emph{weak} and \emph{strong} emergence~\cite{bedauWeakEmergence1997,sethMeasuringAutonomyEmergence2010,yuanEmergenceCausalityComplex2024}. 
Traditionally, weak emergence has been quantified using information-theoretic metrics pertaining to the causal relationship between macroscopic properties and microscopic factors~\cite{crutchfieldCalculiEmergenceComputation1994,sethMeasuringAutonomyEmergence2010,hoelWhenMapBetter2017,rosasReconcilingEmergencesInformationtheoretic2020,greenEmergenceComplexNetworks2023,rodriguez-falconQuantifyingEmergentBehaviors2025}, which rely on probabilistic models of how systems evolve over time. 
Group FeAR, on the other hand, provides a measure of causal responsibility of groups of agents without needing probabilistic or prediction models. 

More assertiveness from other agents mean that the trajectory of the affected agent is less dependent on it's own actions and more dependent on the collective actions of others --- which necessitates coordination among agents. Unlike the case of weak emergence where properties emerge at macroscopic scales~\cite{hoelQuantifyingCausalEmergence2013,hoelWhenMapBetter2017}, the assertiveness of groups of agents act on the trajectories of an affected agent on the same scale which obviates the need to identify the scale at which emergence occurs.
The standard deviation of Kendall's $\tau$ between the rankings of assertiveness from iFeAR and gFeAR are indicative of the superadditivity of assertive influences in a scenario. 
The results from the randomised simulations (\cref{sec:Results-CompareRanks}) showing more group effects for \emph{Aggressive} than \emph{Random}, agree with the intuition that complexity is maximum in the space between complete order and complete chaos~\cite{grassbergerQuantitativeTheorySelfgenerated1986,crutchfieldInferringStatisticalComplexity1989,langtonComputationEdgeChaos1990,mitchellEvolvingCellularAutomata1994,gell-mannInformationMeasuresEffective1996}.
Thus, the standard deviation in Kendall's $\tau$ comparing the ranks of assertiveness provides a model-agnostic metric for detecting the emergence of complexity in spatial interactions.

\subsection{Applications}
Group FeAR along with the tiering algorithm can be used to ascribe backward-looking causal responsibility in spatial interactions. 
As grid-world simulations with discrete actions and non-adaptive agents limit the external validity of our results, future research incorporating continuous formulations of FeAR~\cite{georgeFeasibleActionSpace2025} and agents with explicit normative reflections~\cite{moralesAutomatedSynthesisNormative2013} might be better at explaining empirically observable behaviours in real-life spatial interactions.

As group effects necessitate more coordination between agents, in a decentralised setting, group effects should be minimised.
The presence of group effects might warrant interventions in the form of infrastructural changes (barricades, roundabouts, lanes), active monitoring and policing (crowd control, traffic signals), or through training and testing humans (driver's license).

Furthermore, if we know that a group of agents can communicate or coordinate amongst themselves, then group FeAR can use to generate collective actions that maximise courteousness to other agents by minimising group FeAR.
 
\section{Conclusion}
\label{sec:Conclusion}

We presented a reformulation of the feasible action-space reduction (FeAR) metric to quantify the causal responsibility of groups on the trajectory of an affected agent. Based on marginal changes in FeAR, we identified four types of assertive influences - `solo', `mediated', `coupled' and `mediated coupled'. Base on these assertive influences, we proposed a tiering algorithm for ranking the assertiveness of agents.
Furthermore, through scenario-based simulations, we demonstrated how group FeAR along with the tiering algorithm can be used to identify the emergence of group effects in multi-agent spatial interactions. 
\\
\\
\noindent\textbf{Acknowledgements}: This work is supported by the TU Delft AI Labs programme.

\bibliographystyle{splncs04}
\bibliography{Emergence_and_Responsibility}

\end{document}